\documentclass[a4paper,twocolumn,floatfix,aps,prd,amsmath,amssymb,nofootinbib,preprintnumbers,superscriptaddress]{revtex4}

\usepackage{times,amsfonts}
\usepackage{natbib}
\usepackage[english]{babel}
\usepackage[T1]{fontenc}
\usepackage[latin2]{inputenc}
\usepackage{graphicx}
\usepackage{color}
\usepackage{tabularx}
\usepackage{mathrsfs}
\bibpunct{(}{)}{;}{a}{}{,}
\usepackage{aas_macros}

\usepackage{url} 

\newcommand\tabcaption{\def\@captype{table}\caption}
\makeatother

\newcommand{\incq}[1]{\rm{Q#1}}
\newcommand{\incv}[1]{\rm{V#1}}
\newcommand{\incw}[1]{\rm{W#1}}
\newcommand{\incall}[1]{\rm{INC#1}}
\newcommand{\mm}[1]{\emph{multi-mask#1}}
\newcommand\mmdef{\mm{} }
\newcommand{\regsk}[1]{pixelization scheme#1}
\newcommand{\mmID}[1]{\mmdef\emph{ no. #1} }
\newcommand\kp{\rm{Kp03 }}
\newcommand\Nsim{N_{{\rm{sim}}}}

\newcommand\NsimMCPDFp{N_{\rm{simPDF}(\chisqp)}}
\newcommand\NsimMCPDFq{N_{\rm{simPDF}(\chisqq)}}
\newcommand\NsimCp{N_{\rm{sim}(C_\mathbf{p})}}
\newcommand\NsimCq{N_{\rm{sim}(C_\mathbf{q})}}
\newcommand\Nreg{r}
\newcommand\Nr{N_r}
\newcommand\Nregrm{N_{reg}(\Nreg,m)}
\newcommand\Nmm{N_m}
\newcommand\Npix{N_{pix}}
\newcommand\Npixth{N_{pix th}}
\newcommand\Pth{P_{th}}
\newcommand\chisq{\chi^2}
\newcommand\chisqp{\chisq_{\mathbf{p}}}
\newcommand\chisqq{\chisq_{\mathbf{q}}}

\newcommand\chisqqMC{{\chisqq}_{\tiny{\rm{MC}}}}

\newcommand\Xpi{X_{\mathbf{p},i}}

\newcommand\Xp{\mathbf{X_p}}
\newcommand\XMCmin{{\rm{min}}(\Xpi^{\rm{sim}})}
\newcommand\XMCmax{{\rm{max}}(\Xpi^{\rm{sim}})}
\newcommand\DOFeff{{\rm{DOF_{eff}}}}

\newcommand{\nsigth}[1]{#1 n_{\sigma,th}#1}
\newcommand{\nsig}[1]{#1 n_{\sigma}#1}
\newcommand{\nsigMC}[1]{#1 n_{\sigma}^{\rm{MC}}#1}
\newcommand\secQ{Q_2}
\newcommand\secQpi{{\secQ}_{\mathbf{p},i}}

\newcommand\iCovq{\mathbf{C^{-1}_{\mathbf{q}}}}
\newcommand\vX{\mathbf{X}}
\newcommand\HEALPIX{\emph{HEALPIX }}

\newcommand\highresplotsaddr{}
\newcommand{\pubdataaddr}{\tt http://cosmo.torun.pl/\~{}blew/SKregstat/}

\newcommand{\lb}[2]{$(l,b)=(#1^\circ,#2^\circ)$}
\newcommand{\lmax}{\ell_{\rm{max}}}
\newcommand{\Almax}[1]{A_{\lmax #1}}

\begin{document}
\title{Real space tests of the statistical isotropy and Gaussianity of the WMAP CMB data}
\author{Bartosz Lew} \email[]{blew@a.phys.nagoya-u.ac.jp }

\affiliation{National Astronomical Observatory, 2-21-1 Osawa, Mitaka, Tokyo 181-8588, Japan}
\affiliation{Department of Physics and Astrophysics, Nagoya University, Nagoya 464-8602, Japan}
\affiliation{Toru\'n Centre for Astronomy, Nicolaus Copernicus University, ul. Gagarina 11, 87-100 Toru\'n, Poland}

\begin{abstract}
We introduce and analyze a method for testing statistical isotropy and Gaussianity and apply it to the 
Wilkinson Microwave Anisotropy Probe (WMAP) cosmic microwave background (CMB)
foreground reduced, temperature maps.
We also test cross-channel difference maps to constrain levels of residual foregrounds contamination and systematical 
uncertainties. 
We divide the sky into regions of varying size and shape and measure the first four
moments of the one-point distribution within these regions, and using their simulated spatial distributions we test the 
statistical isotropy and Gaussianity hypotheses.
By randomly varying orientations of these regions, we sample the underlying CMB field in a new  manner, 
that offers a richer exploration of the data  content,
and avoids possible biasing due to a single choice of sky division.
In our analysis we account for all two-point correlations between different regions
and also show the impact on the results when these correlations are neglected.
The statistical significance is assessed via comparison with realistic Monte-Carlo simulations.

We find the three-year WMAP maps to agree well with the isotropic, Gaussian random field simulations 
as probed by regions corresponding to the angular scales ranging from $6^\circ$ to $30^\circ$ at $68\%$ confidence level. 

We report a strong, anomalous ($99.8\%$ CL) dipole ``excess'' in the V band of the three-year WMAP data and also
in the V band of the WMAP five-year data ($99.3\%$ CL).

Using our statistic, we notice the large scale hemispherical power asymmetry, 
and find that it is not highly statistically significant in the WMAP three-year data
($\lesssim 97\%$) at scales $\ell \leq 40$. 
The significance is even smaller if multipoles up to $\ell = 1024$ are considered ($\sim 90\%$ CL).
We give constraints on the amplitude of the previously-proposed CMB dipole modulation field parameter.

We find some hints of foreground contamination in the form of a locally strong, anomalous kurtosis-excess in the Q+V+W co-added map, 
which however is not significant globally. 

We easily detect the residual foregrounds in cross-band difference maps at rms level $\lesssim 7{\rm \mu K}$ 
(at scales $\gtrsim 6^\circ$) and limit the systematical uncertainties to $\lesssim 1.7{\rm \mu K}$ (at scales $\gtrsim 30^\circ$).
\end{abstract}

\accepted{09 July 2008}

\maketitle

\section{Introduction}
\par 

Observational cosmology has established the flat $\Lambda$CDM
model with nearly scale invariant initial density perturbations as the
standard model of modern cosmology
(e.g. \cite{Riess:2004nr,Astier:2005qq,Eisenstein:2005su,Cole:2005sx,Hinshaw:2006ia,Page:2006hz,spergel06,Tegmark:2006az}). 
These observations seem
consistent with the simplest predictions from inflation theory.
Amongst those predictions, one consequence from the cosmological
principle, the statistical isotropy (SI), and one generic consequence from
inflation theories, the Gaussianity (to leading order) of the cosmic microwave background
(CMB) temperature fluctuations,  have received a lot of attention with 
the release of the first year of observations of the WMAP satellite. The
relevant statistical analyses either aimed at detecting small
amounts of non-Gaussianity (NG), that stems from non-linear effect even within
inflation theories \citep{2004PhR...402..103B}, or looked for any
anomalous signal that would challenge this standard model.

However, separating SI from Gaussianity is a delicate task when making
such a test, since one has to deal with only one realization of the
CMB, that is considered in this context to be a random field.
SI and NG have been tested in variety of ways and some ``anomalies'' have
been reported. In particular, using tests optimized for  SI, in
spherical harmonic (SH) phase space \citep{WMAPTegmarkAgainst,dOSS96}
an unusual alignment ($98\%$ CL) at low multipoles have been found and
confirmed (e.g.\cite{2006MNRAS.367...79C,2005PhRvD..72j1302L}). Number
of other tests and statistical tools and estimators have been devised
and used to constrain SI and/or NG.  Among others, these include:
bi-polar power spectrum \citep{2006PhRvD..74l3521H}, 
phase correlations tests \citep{2005PhRvD..72f3512N},  
higher order correlations in SH space (bi/tri-spectrum) e.g.\citep{Ferr98,2004MNRAS.351L...1M,2006MNRAS.tmp..497C,2005MNRAS.358..684C},
$n$-point real space statistics: \citep{durrer-2000-62,2003MNRAS.346...47G,2003PhRvD..68b1302G},
morphological estimators (like Minkowski functionals) \citep{2002MNRAS.331..865S,2001PhRvL..87y1303W,2004MNRAS.349..313P}, 
multipole vectors \citep{2006MNRAS.367...79C,2005MNRAS.362..838L,2004PhRvL..93v1301S,WMAPmultipol}
higher order correlation functions \citep{2003PhRvD..68b1302G},  
phase space statistics \citep{2005PhRvD..72f3512N,2003ApJ...590L..65C},
wavelet space statistics \citep{2006MNRAS.371L..50M,2007ApJ...655...11C,2004ApJ...609...22V,2006MNRAS.369.1858M},
higher criticism statistic \citep{2005MNRAS.362..826C}, 
pair angular separation histograms \citep{PASHtestbyBernui} and also 
various real-space based tests eg: \cite{2004MNRAS.354..641H,2004ApJ...605...14E,2007ApJ...660L..81E,2004ApJ...607L..67H}
In particular a dedicated tests of hemispherical power asymmetry have been reported by many authors and found anomalous at confidence levels ranging from $\sim2\sigma$ to $\sim2.6\sigma$ ($95\% {\rm CL} \sim 99\%$ CL)

In this work, we measure  regional one-point statistics in the WMAP data and in simulations
in order to test the SI and Gaussianity hypotheses.
We mean to extend and generalize the previous similar works in three ways.

Firstly, we show that the result of the analysis strongly depends on the way in which the sky
is partitioned into regions for the subsequent statistics, and we circumvent this problem
by relaxing the constraints on the shape and the orientation of a chosen sky pixelization by considering 
many randomly oriented sky regionalizations. This allows us to avoid a possible bias in  
such regional analysis that is constrained only to a single choice of \regsk{}.

Secondly, we relax the constraint on the size of the regions, thereby statistically probing features at 
different angular scales. 

Thirdly, we account for all correlations between different regions, resulting from the well known two-point correlations
(or possible higher-order correlations) using multivariate full covariance matrix calculus 
for more robust estimation of the statistical significance of local departures from Gaussian random field (GRF) simulations.

We will assess the statistical significance of our results in three
different manners so as to avoid the standard pitfalls of such an
analysis and will rely heavily on realistic simulations to either probe
the underlying distributions or to test the sensitivity of our statistic.

The paper is organized as follows: in Sect.~\ref{sec:data_and_sims} we
introduce the data sets that are being tested, and provide details of
the simulations. In Sect.~\ref{sec:statistics} we describe the details
of our statistical approach for regional statistics. We then test and
illustrate the sensitivity of our statistics  via Gaussian and
non-Gaussian simulations in Sect.~\ref{sec:tests} before presenting 
the results in  Sect.~\ref{sec:results} and discussing them in Sect.~\ref{sec:discussion}. 
We conclude in Sect.~\ref{sec:conclusions}.

\section{Data and simulations}
\label{sec:data_and_sims}

For the main analysis in the paper we use the WMAP three-year foreground reduced temperature maps from differential assemblies (DA) 
Q1, Q2, V1, V2, and W1, W2, W3, W4, pixelized in the \HEALPIX sphere pixelization scheme with resolution 
parameter ${N_s = 512}$. 
We co-add them using inverse noise
pixel weighting (Eq.~\ref{eq:ilc}) and form either individual
frequency combined maps (Q, V, W) or an overall combined map (Q+V+W) to increase the signal to noise ratio according to:
\begin{equation}
T_i = \frac{1}{W_i}\sum\limits_{j=j_{st}}^{j_{en}} w_{ji} T_{ji}
\label{eq:ilc}
\end{equation}
where $W_i = {\sum_{j=j_{st}}^{j_{en}} w_{ji}}$ and  $w_{ji} = N_{ji}/\sigma_{0j}^2$ and ${\sigma_0^{}}$ is the 
noise rms for a given DA and $N_{ij}$ is the number of observations of the $i$th pixel for $j$th DA \citep{2007ApJS..170..288H}.
The sum over $j$ iterates the DAs whose maps are co-added (in numbers, $\{2,2,4,8\}$ respectively for Q, V, W and all channels).
We will refer to those datasets as \incq{}, \incv{}, \incw{} and \incall{ } (inverse noise co-added map) 
respectively and define a data set vector
$d\in\{\incq{}, \incv{}, \incw{}, \incall{}\}$ for further reference.

We also consider a difference maps between different channels to independently test the residual foregrounds and to cross-check 
with the results obtained from the single band NG analysis. We consider a single band difference maps (e.g. Q1-Q2, V1-V2) as well,
since nearly identical frequency difference maps have a negligible amount of CMB or foreground signal\footnote{
The non-vanishing CMB or foregrounds content, even in the single band differential maps, comes from slight 
differences in the effective working frequencies 
of the differential assemblies (DAs) and also from slightly different beam profiles. 
While in case of the single band difference maps (e.g. Q1-Q2) the residual 
rms signal is weaker than the noise by more than two orders of magnitude, 
in case of the different frequency bands (e.g. Q-V) the residual CMB rms 
signal is about one order of magnitude weaker than the noise. }
and these are used to test the consistency of our white noise realizations against the pre-whitened 
$1/f$ pink noise of the WMAP data and constrain the systematical uncertainties. 
Details of this check is given in appendix~\ref{app:q1q2diff_test}.
We will refer to these maps as QV, QW or VW for cross-band difference maps and Q12, V12 etc. 
for an individual differential assembly difference maps.

As an extension to the main analysis we also test the five-year WMAP data set from the V channel and refer to it as \incv{5}.
For this purpose the WMAP five-year simulations are used and preprocessed in the same way as in case of the WMAP three-year
data except for the sky-mask, which here we choose to be KQ75. 

\par The residual monopole, measured outside the three-year release of the Kp0 (hereafter the \kp) sky mask, is
removed from each map by temperature shift in real space. The \kp sky
mask (including galactic region and bright point sources) is applied
and no downgrading is performed at this level. We will use $N_{sim} =
10^4$, realistic, full resolution simulations to test our statistics
and to assess confidence thresholds (see Appendix~\ref{app:simulations} for details and basic tests).

\section{Directional statistics}
\label{sec:statistics}
If the CMB sky is a realization of a multivariate Gaussian random field (GRF), then
statistics of any linear statistical estimator should not deviate from
Gaussianity within any arbitrary region in the sky. Otherwise - in case of non-linear estimators - 
in general deviations from Gaussian statistics are expected, hence MC approach for assessing limits on consistency with
Gaussianity is used. 

In order to test the \emph{stationarity} and Gaussianity  of the temperature fluctuation hypotheses
we use two independent sphere \regsk{s} to define sky divisions and consequently a set of adjacent, continuous regions.

\subsection{Sky pixelizations}
\label{sec:regionalization}

\par
\begin{figure}[!t]
\centering
\renewcommand{\figurename}{Fig}
\begin{tabular}{cc}
{\small {\it HP} 2 } & {\small {\it LB} 32 8}\\
\includegraphics[width=0.239\textwidth]{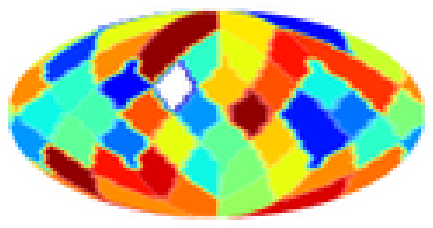}&
\includegraphics[width=0.239\textwidth]{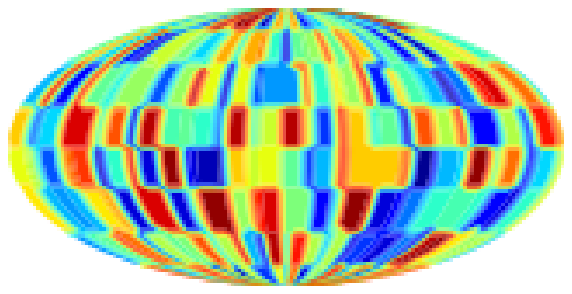}\\
{\small {\it HP} 2 } & {\small {\it LB} 64 8}\\
\includegraphics[width=0.239\textwidth]{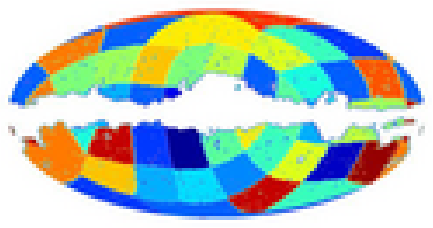}&
\includegraphics[width=0.239\textwidth]{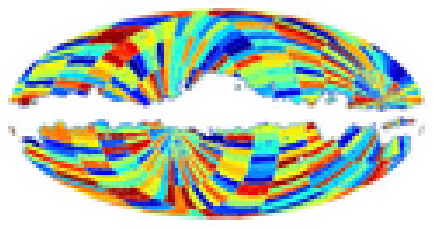}
\end{tabular}
\caption{In the first row, two lowest-resolution pixelization schemes -- {\it HP} 2
  (top-left) and {\it LB} 32 8 (top-right) are shown.
In the second row, we present an examples of two \mm{s} actually used
  in analysis. Pixelization schemes are rotated to a random orientation, with \kp sky mask applied. These are {\it HP} 2 (lower-left)
  and {\it LB} 64 8 (lower-right) respectively. Values in all regions were randomized for better visualization.
\highresplotsaddr
}
\label{fig:regionalizations}
\end{figure}

 The first \regsk{} (hereafter referred to as \mbox{{\it HP})} is an independent implementation of the \HEALPIX pixelization
scheme \citep{Healpix2005} and its resolution is parametrized by the
$n_s$ parameter (Fig.~\ref{fig:regionalizations}
top-left). The total number of pixels for a given $n_s$
is $\Nreg = 12 n_s^2$. We will use three different
resolutions ($n_s$) as specified in Table~\ref{tab:regionalizations}.

\par The second one (hereafter called {\it LB}) covers the sphere by dividing it
along lines of parallels (iso-latitude) and meridians (iso-longitude) to
obtain arbitrarily elongated pixels, generally of varying angular sizes (Fig.~\ref{fig:regionalizations}
top-right). This results in the total
number of pixels $\Nreg = N_l N_b$ (where $N_l$ and $N_b$ are the
numbers of longitudinal and latitudinal divisions). The three
different resolutions used in the analysis defined by these parameters are specified in
Table~\ref{tab:regionalizations}. Further flexibility is allowed by
rotating the polar axis by three randomly chosen Euler angles. 

\begin{figure}[!t]
\centering
\renewcommand{\figurename}{Fig}
\includegraphics[width=0.49\textwidth]{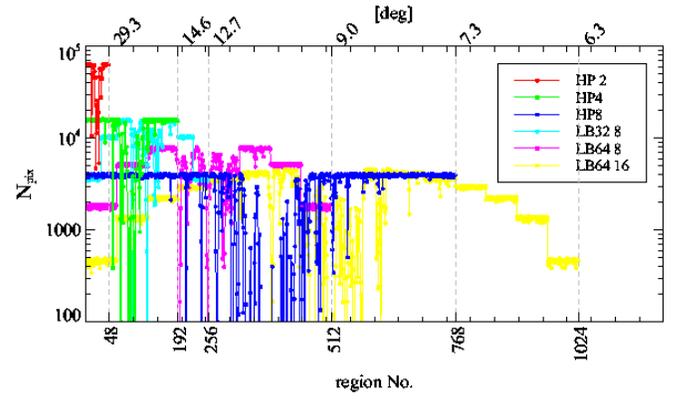}
\caption{Number of points, $\Npix (k)$, in regions for all 6 types of \regsk{s } used in their initial position, 
after masking by \kp sky mask, in function of region number. All regions with $\Npix < 100$ were not
considered in the analysis as detailed in the text and treated as masked. The
central parts ($\sim \Nreg/2$) of a given pixelization are
strongly covered by the \kp mask (this is true only for the particular initial orientation
of  a \mm{}). At the top abscissa we give the approximate angular scales probed by \regsk{s} with the corresponding
total number of regions indicated in the bottom abscissa.}  
\label{fig:point_num-all}
\end{figure}

Since there is no reasonable, physical motivation for preferring any particular
sky pixelization over another from the standpoint of testing a GRF hypothesis, 
we consider $\Nmm=100$ random orientations 
for each of the six types of \regsk{s}, which altogether yields 600 different sky pixelizations 
with a total number of $280\,000$ regions of
different shapes and sizes probing different angular scales (Fig.~\ref{fig:point_num-all}).  
We therefore draw the three Euler angles used to define the axis position and \regsk{} orientation about this axis 
from a uniform distribution.

All sky pixelizations are subject to the \kp (three-year Kp0) galactic/point sources cut which masks $\sim 23\%$ of the sky.
In practice there is no lower bound for the size of a region due to its random orientation with respect to the \kp sky cut.
However for the sake of numerical stability, when computing the inverse
covariance matrix (see below), we only consider regions that happen to
have $\Npix > \Npixth = 100$, where $\Npix$ refers to the number of
pixels of the original $n_s=512$ map falling into this particular region. 

Hereafter we refer to a particular random realization of a \regsk{} (a random set of
regions covering the full sky and merged with \kp sky mask) as a \mm{}, since it uniquely tags
sky regions and allows to pursue statistics exclusively within them
(see Fig.~\ref{fig:regionalizations} bottom-left and bottom-right). Of
course different \mm{s}, even defined from a similar \regsk{}, may have
a different number of regions due to the random orientations with respect to the
\kp sky mask. We define $\Nregrm$ as the number of regions of a
\mm{} as a function of initial resolution parameter $r$ and \mm{} ID number
$m\in\{1..\Nmm\}$. 
As an illustration, the two lowest resolution, \regsk{s } and two examples of \mm{s} are shown in Fig.~\ref{fig:regionalizations}.
We will also use additional sets of \mm{s} to complete and extend the main part of the analysis in a few selected cases.
\begin{table}[!b]
\caption{Summary on the {\it LB} and {\it HP} \regsk{s} and resolutions used in the main analysis, given explicitly for quick reference.
The columns abbreviations are as follows: (1) \regsk{} reference name, (2) resolution parameter value, (3) approximated 
angular size of regions, (4) number of regions in \regsk{}.}
\centering
\begin{footnotesize}
\begin{tabular}{llllllllll}\hline\hline
\multicolumn{4}{c}{{\it HP}}&\multicolumn{6}{c}{{\it LB}}\\
(1) & (2)& (3) & (4) & (1) & \multicolumn{2}{l}{(1)} & \multicolumn{2}{l}{(3)} & (4) \\
Ref. & Res.  & Ang.size [deg] & regs. & Ref. & \multicolumn{2}{l}{Res.}  & \multicolumn{2}{l}{Ang.size [deg]} & regs. \\
name&$n_s$  & $\Omega_{\rm reg}$ & $\Nreg$ & name & $N_l$  & $N_b$ & $\Delta_l$ &$\Delta_b$ & $\Nreg$\\\hline
{\it HP} 2&2  & 29.3 & 48  & {\it LB} 32 8  & 32 & 8 & 11.3 & 22.5 & 256\\
{\it HP} 4&4  & 14.6 & 192 & {\it LB} 64 8  & 64 & 8 & 5.6 & 22.5 & 512\\
{\it HP} 8&8  & 7.3  & 768 & {\it LB} 64 16 & 64 & 16 & 5.6 & 11.3 & 1024\\\hline
\end{tabular}
\end{footnotesize}
\label{tab:regionalizations}
\vspace{-0.5cm}
\end{table}

\subsection{One-point statistics}
\label{sec:One-point statistics}

In each of the defined regions of each \mm{}, the first four central moments (i.e. mean (m),
standard deviation ($\sigma$), skewness (S), and kurtosis (K)), of the underlying temperature
fluctuations are computed for the data and for all $\Nsim=10^4$ simulations. 
Together this yields $2.8\times 10^9$ regions for assessment of uncertainties. As will be shown in the
Sect.~\ref{sec:tests}, allowing for arbitrary orientations of \mm{s } has
an impact on the results and yield a more stringent test on
stationarity.
The fact that we choose to work in real space allows for a good localization of deviations in the sky. 

The presence of extended, residual foregrounds or unremoved, unresolved point sources, will
affect the local central moments distributions. 
In particular the mean
of the fluctuations will tend to be up-shifted with respect to simulations if
diffuse foregrounds are present or down-shifted if they are
over-subtracted. Also, depending on the amplitude of the residual
foregrounds the local variance will also be altered. Looking jointly
at the distribution of these moments on large scales might also provide
a handle on the large scale distribution of power via the off-diagonal terms of the inverse covariance matrix.
The physical extent and position of the regions where particular type of deviation occurred 
can provide a clue to the possible nature of the foregrounds causing it (see Sect.~\ref{sec:tests}). 

\subsection{Assessing statistical significance}
\label{sec:statistical-approach}

Since our measurements are statistical, a crucial stage
remains in probing  their statistical significance. 
Our approach relies on a detailed comparison between the measurements
performed on real data with the distribution of the same measurements performed on
simulations. 

We consider three different ways to address the significance of these
measurements. Each step involves one extra-level of generality and
will shed light on the subtleties of such an assessment.

At first, we look at individual regions, ignore their correlations and compare them with the simulations. We call
this approach the ``individual region analysis''. It is the simplest approach one can consider.

Secondly, we compute the overall statistical significance per \mm{}, by taking into account
the two-point correlations between moments of distributions (MODs) measured in regions of the same \mm{} via the full covariance
matrix. We call this approach the ``multi-region analysis''. The resulting
probabilities $P\bigl(\chisqq\bigr)$ (Eq.~\ref{eq:Pjoint}) are the joint probabilities of exceeding a certain
confidence threshold as a function of \regsk{} ($r$), \mm{} ($m\in\{1..\Nmm\}$), MOD ($X\in\{\rm{m},\sigma,\rm{S},\rm{K}\}$), 
and dataset ($d\in\{\incq{}, \incv{}, \incw{}, \incall{}\}$) configured by a parameter vector $\mathbf{q}=\{X,r,m,d\}$.
This analysis extends the information from the single region analysis by
testing the consistency of the data with the simulations via standard
multivariate calculus.

Finally, we combine all the information probed by different \mm{s} to find the 
joint cumulative probability of rejecting the GRF hypothesis as a function of 
\regsk{} ($r$), MOD ($X$), and dataset (eg. frequency) ($d$). 
We call this approach the ``all \mm{s} analysis''.

We remind that the statistical significance of any real data measurement, at any stage of the analysis,
is always assessed by a comparison to the set of the same measurements performed using GRF simulations.
The exact details of the analysis at each step are given in Appendix~\ref{app:analysis}.

\subsection{Visualizing the results}
\label{sec:visualizing_results}

To visualize our results from the single-region analysis, or multi-region analysis at certain confidence level, 
we proceed the following way. For individual region statistics, for
each region of each \mm{} we define $\nsig{}$ as   
\begin{equation}
n_\sigma = \sqrt{2}{\rm{erf}}^{-1}(1-P(X)) = {\rm{cdf_G}}^{-1}(P(X)/2)
\label{eq:nsigma}
\end{equation}
where $P(X)$ is the quantile probability derived according to Eqs.~\ref{eq:interpolation} and~\ref{eq:extrapolation}.
The $\nsig{}$ thus defined is the Gaussian number of $\sigma$s by which a region, defined by a given
\mm{}, deviates from simulation average. We then produce maps of $n_\sigma$ estimator, for data processed
through each of the 600 generated \mm{s} and for each MOD. Then, to
present all the results in a compact way, we scramble these maps within the same MOD. 
We over-plot the individual pixels from regions with the strongest deviations from the underlying pixels. Positive
$\nsig{}$ values correspond to excessive value of a given MOD in a
region, and negative value correspond to its suppression. For clarity,
we use a threshold $\nsigth{|} = 3$ to produce maps with only the
strongest ($3\sigma$) detections. 

For the joint multi-region statistics we produce maps (as detailed
above) using only those \mm{s} that yield $|P(\chisqq)| \leq \Pth$
(Eqs.~\ref{eq:interpolation},~\ref{eq:extrapolation}) revealing
detections at the statistical significance $1-\Pth$ for a given MOD, \mm{}
resolution $\Nreg$ and dataset $d$, i.e. for a given parameter
$\mathbf{q}$. Within this notation, single region statistics
corresponds to $\Pth=1$. 

While the ``n-sigma'' maps are easy to read when looking at distributions of the anomalies at a given local 
significance, they are unitless and cannot be directly linked with quantities that are physically measured.
We therefore also consider a difference maps ($\Delta$ maps\footnote{We will hereafter refer to maps so produced
as $\Delta$ maps to make a clear distinction from the difference maps obtained by differentiation of temperature
maps from different frequencies (e.g. QV, QW, etc...)}) of regional departures in individual MODs between datasets and 
averaged simulation expectation: i.e. for $i$th region of a given \mm{} we plot $\Delta_i=X_i-\langle X_i\rangle_N$,
where $\langle\rangle_N$ stands for average over N simulations.

\section{Tests of the simulation and measurement pipeline}
\label{sec:tests}

In order to validate the statistical tools introduced above, 
test the sensitivity and the correctness of the numerical code, 
we performed a set of
experiments using both simulated WMAP CMB data and data with either simulated
violation of the large scale statistical isotropy or localized NG
features. 

\subsection{Consistency check with GRF simulations}
\label{sec:consistency_check}

We first test self-consistency by generating 10 additional \incall{} CMB data sets 
and carry out for each of them the single-region, joint multi-region and all \mm{s}
statistics. As expected, we found that the simulated datasets yield a
good consistency with the simulations (at $68\%$ CL).

\subsection{Sensitivity to local NG}
\label{sec:sensitivity_to_NG}

For a statistically isotropic Gaussian process the
kurtosis is expected to be exactly equal to ${\rm K}=3$, which translates
into a kurtosis excess ${\rm KE}=K-3=0$. 
Violation of either of the assumptions can lead to a positive or negative KE.

At first we simulate what could be a residual component, resulting from subtraction of an non-ideal foreground
template,  extended
over an area of $10^\circ$ angular, centered at \lb{50}{50}, whose
signature would be a non-vanishing KE. 
Such residuals must be expected to be small in the foreground-cleaned maps, and they could be either positive -- resulting from 
foregrounds undersubtraction -- or negative ones -- resulting from foregrounds oversubtraction.

Therefore, the introduced NG component
is drawn from a normal
distribution with variance $\sigma^2$ and with a mean increasing
uniformly across the patch when going north-south following Healpix
ring ordering inside the spot. From the northern point of the patch down to the
southern part of the patch, the mean will shift by 2$n\sigma_{\rm CMB}$.

\begin{figure}[!hbt]
\centering
\renewcommand{\figurename}{Fig}
\includegraphics[width=0.5\textwidth]{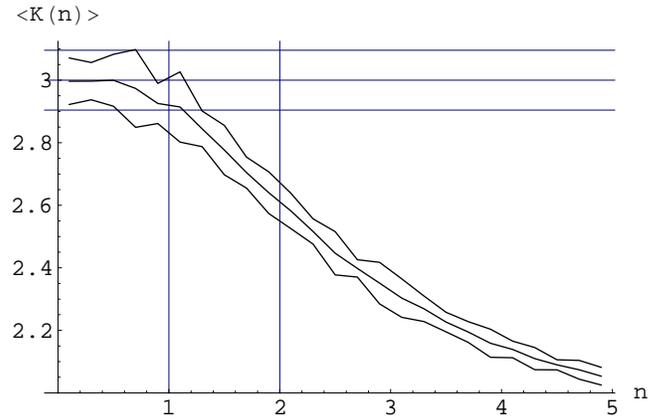}
\caption{Negative departure of the kurtosis in the NG spot as a function of $n$ parameter. 
The plot shows the kurtosis of a sum of GRF and the generated 
NG template averaged over 10 random realizations of the two. The variance of the 
GRF is 100 larger than the variance of the NG template and the field size is chosen to correspond to the NG spot size as described in the text. 
The $3\sigma$ confidence thresholds were quoted around the expected values for the GRF only and for the GRF+NG template cases.
Vertical lines indicate the chosen values of the $n$ parameter used for the templated maps generation
for the test.
}
\label{fig:Kavvsn}
\end{figure}

We introduce in this way a gradient in the noise and thus a non-zero local negative KE (Fig.~\ref{fig:Kavvsn}) but
still preserve a vanishing (within the spot) skewness. The $\sigma$ value is chosen to be
1\% of the underlying CMB rms ($\sigma_{\rm CMB}$). 
To test the sensitivity of our estimator we consider $n$ to be either 1 or 2, above which 
the NG template starts to be visually noticeable due to edge discontinuities. 
Note that, with a so defined anomaly, the pixels of the spot that are close to its horizontal diameter
will have the least impact on the underlying CMB field distortions.

The choice of the $n$ parameter values corresponds to the NG signals of the rms amplitude $\sim 50{\rm \mu K}$
and $\sim 100{\rm \mu K}$ for $n=1$ and $2$ respectively within the spot.
We note that the rms. value within the spot of the same size, in the foregrounds reduced difference VW map of the WMAP3 data,
yields $> 100{\rm \mu K}$ depending on the exact location of the spot in the sky, hence our choice of NG signals amplitude aim at detection 
of relatively small anomalies as compared to the WMAP3 noise specifications.

We find that for a single NG spot of radius $10^\circ$, the multi-region analysis does
not return any significant detection for $n=1$ in any of the MODs, but the single region analysis
finds the contaminated regions unusual at $\nsig{}\approx 2.5$.
In case of $n=2$ we detected a $3\sigma$ local deviation in kurtosis, while in all-\mm{s} analysis
we reject Gaussianity at $99.8\%$ CL ({\it HP} 8) due to variance distributions\footnote{Of course, 
the estimated rejection confidence thresholds given in Table.~\ref{tab:all-mm-joint-tests} based only on a 
single-simulation measurements may be somewhat biased depending on particular realization of the GRF simulation.}.

\begin{figure}[!t]
\centering
\renewcommand{\figurename}{Fig}
\begin{tabular}{ll}
 {\small a) template} & {\small b) variance} \\
\includegraphics[width=0.239\textwidth]{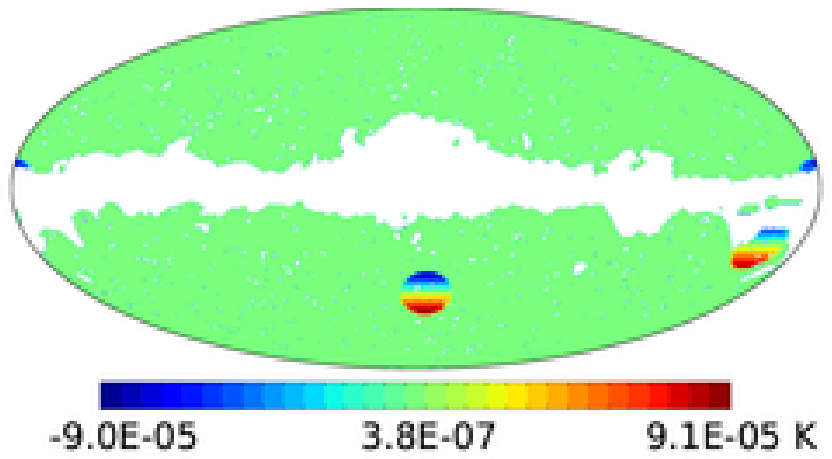}&
\includegraphics[width=0.239\textwidth]{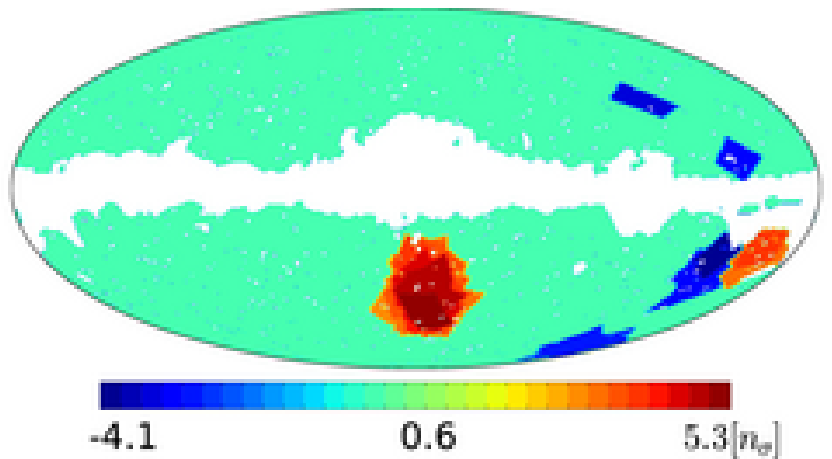}\\
 {\small c) skewness} & {\small d) kurtosis} \\
\includegraphics[width=0.239\textwidth]{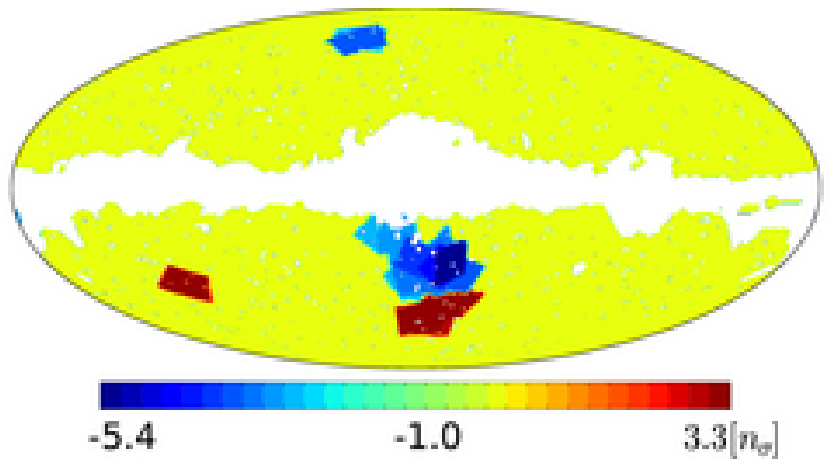}&
\includegraphics[width=0.239\textwidth]{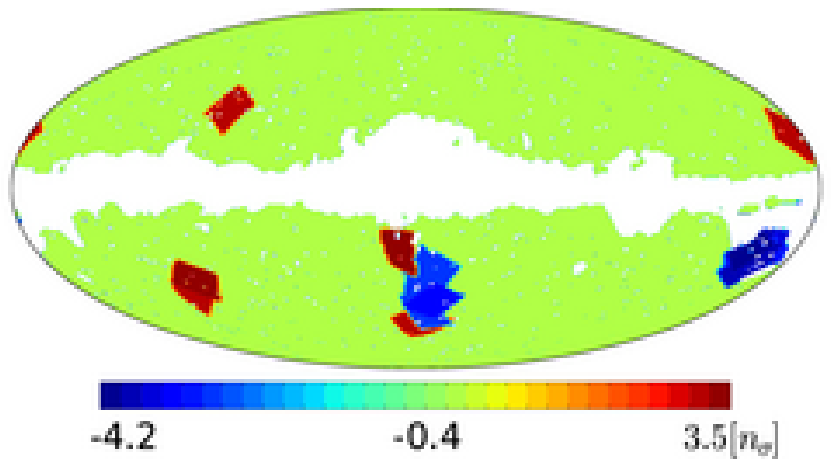}\\
\end{tabular}
\caption{{\it a)} A Non-Gaussian temperature gradient template, leading to locally negative kurtosis excess. 
The parameter value of $n=2$ is used (see the main text for details). Note that one of the spots is practically removed by the galactic sky cut. 
{\it  b-d)} The results of the single-region analysis using the 
100 \mm{s} of the {\it HP} 4 \regsk{} of one of the simulated INC data and contaminated with the template.
The scrambled $\nsig{}$ detection map in variance, skewness and kurtosis, thresholded at $\nsigth{}=3$, is plotted.
Note the strong, template-induced, local anomalies detections traced by different \mm{s}, as well as some other, but somewhat weaker, 
detections of $\nsig{|}>3$ regions. 
In particular, the template leads to the local kurtosis suppression and strong local excess of the variance.
}
\label{fig:3NGKneg_template}
\end{figure}

We rerun the test for the same type of NG templates but replicated in 3 disjoint spots at
different directions in the sky for the same values of $n$
parameter (Fig.~\ref{fig:3NGKneg_template}-a). 

The choice of the position and the size of the spots is most relevant to the results presented in Section~\ref{sec:results}.
The results of the single region analysis is shown in Fig.~\ref{fig:3NGKneg_template}(b)-(d).
Note how different \mm{s} trace the locally introduced anomaly. Depending on the orientation of the \mm{} and its 
regions around the directions of the NG spots, the returned $\nsig{}$ values differ. In the overall multi-region analysis
this naturally leads to a distribution of probabilities which strongly depend on how the features of the map are split and probed 
by different regions.

Note that some of the \mm{s} also return an $\nsig{}>3$ detections even in a template-free regions.
It is therefore clear that use of many differently oriented \mm{s} helps to investigate the statistical significance
of local anomalies.

The multi-region analysis in case of $n=1$ return no significant detections in any of the MODs,
but very significant deviations were detected for the case $n=2$ in
all-\mm{s} analysis (Table~\ref{tab:all-mm-joint-tests} in section "KE-"), again only in the variance distributions.

Consequently, we find that for the unfiltered maps the distributions of variances are actually more sensitive to this kind of
simulated anomalies, rather than higher order MODs. We note, however,  that measuring a local sign of KE may be a hint
of the nature of the foregrounds signals as the kind of template used in this example introduces 
locally only the negative KE as shown in Fig.~\ref{fig:Kavvsn}.
Similar dependences are obtained for templates 
of different shapes and sizes and combinations of these factors. 
In the limit of a flat template ($n=0$) the field becomes Gaussian, as expected.

It is possible to introduce a non-vanishing skewness by making the template unsymmetrical,
by considering shifts in the mean which are unsymmetrical about zero. A similar effect is possible by 
considering regions of \mm{s} that only partially
overlap with the area of the NG spot (Fig.~\ref{fig:3NGKneg_template}c).
Note that the example from Fig.~\ref{fig:Kavvsn} does not include the effects of the non-uniform
noise component which is included in our tests and also that the confidence level contours
were derived assuming the Gaussian error statistics; therefore
it is not straightforward to extrapolate the strong detections expected from Fig.~\ref{fig:Kavvsn}  for $n=2$
onto the full sky, locally templated, signal and noise realizations, subject to subsequent regional statistics.

Note that the assumed size of local deviations is small relative to the full sky one, and hence their global impact 
is reduced accordingly. Larger, in sense of area, deviations will be of course easier to detect.
Also a specific pre-filtering in the spherical harmonic (SH) space, of the data prior to the test may help to
expose the most relevant scales to the test. In this test we focused on testing unfiltered maps; therefore, necessarily 
the strength of the detections must be suppressed.

It is not possible to obtain a locally positive KE with the above-described template. 
Such deviation could however be
the signature of the unresolved point source contribution\footnote{Although this 
would have a specific frequency dependence, we ignore this fact here.},
or an unknown and localized noise contribution.

Again, for qualitative studies only, we simulate the point source component in the full sky by adding random numbers
drawn from a distribution which is an 
absolute value of the Gaussian distribution with zero mean and random variance $n \sigma_{\rm CMB}$ parametrized by parameter $n$ and
uniformly distributed within range $[0,n \sigma_{\rm CMB}]$, where $\sigma_{\rm CMB}$ is the rms value of the underlying CMB fluctuations.

From our simulations however, it appears that it is difficult to detect a significant contribution due
to point source contamination since such signal is significantly smeared by the instrumental beam
even for $n$ as large as 6. Even when  ${\rm KE} \gtrsim 6$ before beam
smearing, the variance response is much stronger than the KE response
leading to inconsistencies with simulations in the total power of the
map as measured by e.g. full sky variance distribution (Fig.~\ref{fig:simshists}). 
We therefore conclude that it is unlikely
to detect any point source contribution to KE in this test which is
not surprising since we work at fairly low resolution which dilutes the
point source signal. 

As already mentioned, a locally generated noise-like component in the map, in principle, could generate an non-negligible positive KE\footnote{
This is most easily seen in the contribution of the anisotropic noise of the WMAP to the kurtosis of the signal only simulated map,
which induces a significant positive overall KE value in the simulation.}
as it would not be processed by the instrumental beams. 
However since such noise is not well motivated physically due to non-local properties of the TOD data and scanning strategy of the WMAP,
and also since the noise properties are well constrained, therefore we do not consider such case.

We conclude that small and single (compared to full sky observations) localized NG features will be difficult to detect 
via higher order MODs in the joint multi-region and all-\mm{s}
analysis due to their small statistical impact on the overall
statistics. However a single region statistics carried out first might
be a rough guide in selecting a possibly interesting foreground NG signals. 
If these indicate regions with significant deviations in variance and possessing a negative KE
it would hint on residual, large scale foreground contamination.

\subsubsection{Stability of results in function number  of \mm{s}}
As different \mm{s} probe the underlying data differently, the joint-probabilities per \mm{} differ
and lead to a distribution that typically covers a wide range of possible probability values.
As such, the multi-region analysis (see sec.~\ref{sec:multi_region} for details)  can be used to find the orientation of the most unusual regions
in the \mm{s} that yield the smallest probability as comparred to GRF simulations.

Since from the point of view of statistical isotropy all \mm{s} are equivalent, in the all-\mm{s} analysis (see sec.~\ref{sec:all_joint} for details)
we integrate the infomation from all \mm{s} within a \regsk{} to obtain an average level of consistency of the data with GRF simulations
 for that  \regsk{}.
This approach also provides a conservative way of averaging over a possibly-spurious detections that could be a fluke,
due to some accidental arrengement between a \mm{s} and a data set.
If the anomalous feature in the map is strong enough to be detected in many \mm{s} then this will also result
in a detection in the joint all-\mm{s} analysis. Conversly, if only one or few \mm{s} result in a very small probability 
the overall impact will not be large due to stability of the median estimator with respect to the distribution outliers.
However, the investigation of the most anomalous \mm{s} may help in selecting the deviating regions for further analyses.

In this section we show how the convergence to the results of the all-\mm{s} analysis is reached in function of 
number of \mm{s} used to derive the median $\chisq$ value and the corresponding median $\chisq$-distribution leading to 
the all-\mm{s} probability.
\begin{figure}[!t]
\centering
\renewcommand{\figurename}{Fig}
\includegraphics[width=0.49\textwidth]{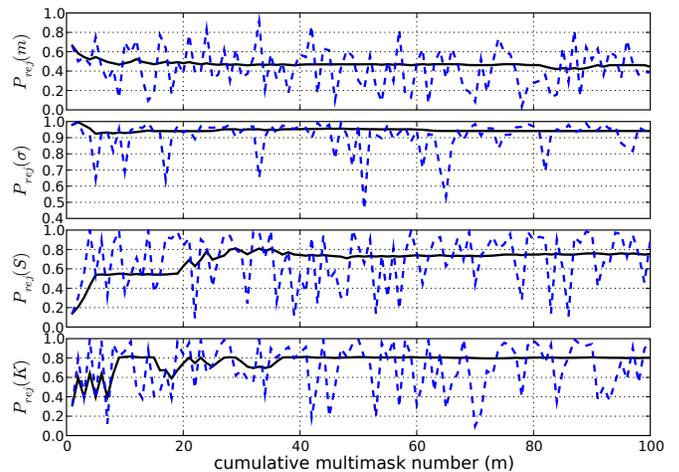}
\caption{Convergence of the joint all \mm{s} probability to the value reported in Table~\ref{tab:all-mm-joint-tests} (in section (2KE-) for $n=2$) 
for the case of {\it HP} 2 \regsk{} in function of the cumulative number of \mm{s} used to derive it (black solid lines) for mean (top panel), variance, skewness and 
kurtosis (bottom panel). Additionally we overplot the joint multi-region probabilities per \mm{} for each \mm{} number (blue dashed lines).}
\label{fig:convergence-Pjointall-vs-mmno}
\end{figure}
As shown in Fig.~\ref{fig:convergence-Pjointall-vs-mmno} the convergence of the joint all-\mm{s} probability to the reported value for $\Nmm=100$,
is in this case ({\it HP} 2 \regsk{}) quite fast; however in general it depends on particular properties of the map as well as on the set and ordering of the \mm{s} used.
The speed of the convergence and the robustness of the final value is as good as the convergence of the unbiased mean estimator - i.e. the median - 
to the intrinsic mean value as the number of random trials (corresponding to the number of \mm{s}) probing the underlying distribution increases.

Note how strongly the joint probability per \mm{} depends on the orientation of the \mm{} (blue-dashed lines in Fig.~\ref{fig:convergence-Pjointall-vs-mmno}).
In particular for the variance case, the probabilities of rejection per \mm{} range from $45.7\%$ for \mm{} $m=51$, to 99.8\% for \mm{} $m=2$.
Yet the, reported in all-\mm{s} analysis, median value is very stable with respect to these variations.

\subsubsection{The shape of \mm{} regions.}
Naturally, the \mm{s} having large regions will be insensitive to the small scale map features,
while the \mm{s} having small regions will be insensitive to the large scale structures.
This motivates the usage different number of regions to probe different scales of the map.

Theoretically there are infinitely many ways of defining the shape of regions of \mm{s},
and of course, our choice of the shape of \regsk{s} and the \mm{s} is somewhat arbitrary; however the 
motivation for using different shapes of regions is straightforward: to probe the data using different binning techniques
since the GRF statistically should not depend on it. However if the data turns out to be non-random
it is possible that the non-randomness will be explored differently by different region shapes.
A loose analogy to the real-space \mm{} region shape, 
which is used to derive a local value of an estimator, over the defined area
for a given orientation of \mm{}, is in the wavelet space the shape of the mother wavelet, 
which is used to obtain the local convolution coefficients for the input map.
In this analogy the size of the region correspons to the wavelet scale.

\subsection{Sensitivity to the large scale phase anomalies}
\label{sec:sensitivity_to_phase_anomalies}

We test the sensitivity of our method to the well known large scale anomalies found in the WMAP data: i.e
to the aligned and planar low multipoles $\ell=2$
and $\ell=3$. In order to test such anomalies we generate two GRF CMB
simulations. In both simulations we use the same realization of the
power spectrum and phases as in the case of the first GRF simulation (Sect.~\ref{sec:consistency_check}).

In the first simulation we enforce large scale phase correlation by introducing  an
``m-preference'' in the power distribution in the quadrupole ($\ell=2$)
and octupole ($\ell=3$). We choose the ``sectoral'' spherical
harmonic coefficient $a_{ll}$,  to carry all the power of the multipoles. In
the second simulation we extend this modification up to $\ell=5$. The
GRF signal simulations are rotated to a preferred frame prior the
introduction of the planar multipoles. The signal maps are then
rotated back to the original orientation so that the maximal momentum
axis was located at \lb{260}{60} before adding noise. 

As shown in Table~\ref{tab:all-mm-joint-tests} such anomalies have not been significantly detected at tested scales.
Although it is expected and observed that considering unfiltered maps (containing 
all multipoles information mixed together) there will be a little overall impact on the statistics we note
a higher sensitivity would be obtained is a prefiltered in SH space data were used.

\subsection{Sensitivity to the large scale power anomalies}
\label{sec:sensitivity_to_power_anomalies}

We give a special attention to testing the sensitivity of the method for detecting and quantifying the previously
reported large scale anomaly in the power distribution in the sky \citep{2004ApJ...605...14E,2007ApJ...660L..81E}. 
We create a simulated CMB maps where the CMB signal is modulated according to:
\begin{equation}
\begin{array}{lll}
T(\mathbf{\hat n}) &=& T_{CMB}(\mathbf{\hat n}) \bigl(1+ M(\mathbf{\hat n})\bigr)\\
M(\mathbf{\hat n}) &=& \Almax{} \mathbf{\hat n}\cdot \mathbf{\hat d}
\end{array}
\label{eq:power_mod}
\end{equation}
where $\mathbf{\hat n}$ is a unit vector and $M$ is a bipolar modulation field,
oriented in direction $\mathbf{\hat d}=(225^\circ,-27^\circ)$ with amplitude 
$\Almax{}\in\{0.114,0.2\}$ which modulates the CMB component up to the maximal multipole of $\lmax=1024$.
\begin{table}[!t]
\caption{All-\mm{s} analysis results from tests of the simulated datasets.
The columns content is as follows: (1) \regsk{}, (2..4) ``probability of rejecting'' the consistency with GRF simulations
(Sect.~\ref{sec:all_joint}) for each MOD.
The abbreviations of the datasets are: 
1KE-: \incall{} simulation with one NG spot (three NG spots) leading to KE$<0$ and  $n=1$ (see text Sect.~\ref{sec:tests} for details),
2KE-: \incall{} simulation with one NG spot (three NG spots) leading to KE$<0$ and  $n=2$ (see text Sect.~\ref{sec:tests} for details),
A2..3: \incall{} simulation with aligned multipoles $\ell=2$ and $\ell=3$,
A2..5: \incall{} simulation with aligned multipoles from $\ell=2$ to $\ell=5$,
M: \incall{} dipole modulated simulation with dipole amplitude of $\Almax{}$. 
<M>: \incv{} simulation with CMB signal fully (partially) modulated according to parameter $\Almax{}$. 
The average confidence thresholds are given.
The values are rounded to integer percentiles in case of probabilities $\leq 99\%$ CL.
The saturated values are marked with $\star$.
}
\centering
\begin{footnotesize}
\begin{tabular}{llllll}\hline\hline
 Reg.sk.\hspace{.3cm} & $P_{rej}^m$[\%]\hspace{.3cm} & $P_{rej}^\sigma$[\%]\hspace{.3cm} & $P_{rej}^S$[\%]\hspace{.3cm} & $P_{rej}^K$[\%]\hspace{.3cm} \\
(1) & (2) & (3) & (4) & (4) \\\hline
\multicolumn{5}{c}{\underline{$n=1$: 1KE- (1KE-):} }\\
&\multicolumn{4}{c}{no significant detections (no significant detections)}\\\\
\multicolumn{5}{c}{\underline{$n=2$: 2KE- (2KE-):} }\\
  {\it HP} 2 & 42 (45)  & 82 (94)  & 58 (75)  & 49 (80)  \\
  {\it HP} 4 & 43 (69)  & 97 ($>99.98$)  & 31 (66)  & 39 (62)  \\
  {\it HP} 8 & 8 (70)  & 99.8 (99.8)  & 5 (32)  & 23 (41)  \\
  {\it LB} 32 8 & 28 (58)  & 88 (99.0)  & 29 (72)  & 27 (53)  \\
  {\it LB} 64 8 & 34 (64)  & 69 (96)  & 20 (60)  & 18 (34)  \\
  {\it LB} 64 16 & 18 (69)  & 98 (99.5)  & 10 (37)  & 28 (44)  \\
\multicolumn{5}{c}{\underline{A2..3 (A2..5): }}\\
&\multicolumn{4}{c}{no significant detections (no significant detections)}\\\\
\multicolumn{5}{c}{\underline{M: $A_{1024}=0.114$ ($A_{1024}=0.2$)} }\\
   {\it HP} 2 & 39 (36)  & 94 (99.99)  & 42 (50)  & 32 (35)  \\
   {\it HP} 4 & 43 (47)  & 99.6 ($>99.99^\star$)  & 22 (25)  & 31 (32)  \\
   {\it HP} 8 & 11 (19)  & 99.96 ($>99.99^\star$)  & 7 (8)  & 18 (20)  \\
   {\it LB} 32 8 & 30 (35)  & 97 ($>99.99^\star$)  & 20 (22)  & 22 (23)  \\
   {\it LB} 64 8 & 34 (38)  & 91 ($>99.99^\star$)  & 15 (17)  & 14 (14)  \\
   {\it LB} 64 16 & 22 (30)  & 99.4 ($>99.99^\star$)  & 12 (13)  & 23 (25)  \\\\
\multicolumn{5}{c}{\underline{<M>: $A_{1024}=0.114$ ($A_{40}=0.114$)} }\\
  {\it HP} 2 & 58 (59)  & 99 (56)  & 52 (44)  & 54 (44)  \\
  {\it HP} 4 & 77 (56)  & 99.87 (57)  & 54 (37)  & 50 (38)  \\
  {\it HP} 8 & 79 (62)  & 99.97 (55)  & 64 (29)  & 54 (52)  \\
  {\it LB} 32 8 & 78 (57)  & 99 (55)  & 54 (38)  & 51 (47)  \\
  {\it LB} 64 8 & 78 (58)  & 97 (55)  & 59 (38)  & 54 (43)  \\
  {\it LB} 64 16 & 76 (61)  & 99.7 (54)  & 64 (32)  & 55 (46) 
\end{tabular}
\end{footnotesize}
\label{tab:all-mm-joint-tests}
\end{table}

The result of the test with such modulated simulation is given in Table~\ref{tab:all-mm-joint-tests} (in section ``M'').
As the amplitude of $A=0.114$ has been previously claimed to be preferred for the CMB data
\citep{2007ApJ...660L..81E} we process five additional, full resolution \incv{} simulations, modulated with this amplitude, 
to reduce a potential biases from a single draw of a random simulation, and report (Table~\ref{tab:all-mm-joint-tests} in section ``<M>'') 
the average rejection thresholds as a function of the \regsk{}.

We also process five additional modulated simulations, and apply the modulation only to the range of multipoles
$\ell \leq 40$, leaving higher multipoles unmodulated, since the aforementioned work operated at much lower resolution.

The test is able to reject the modulation of the CMB of the amplitude $A_{1024}=0.114$ at a very high confidence level
($99.9\%$) depending on the \regsk{}. Note however, that this model modulates all scales equally.

Although in principle, the modulation  will change the underlying power spectrum at scales where it was applied, 
we estimate (Appendix~\ref{app:modulation_power_change_test}) that any such effect for the modulation of $A_{1024}=0.114$ still remains 
in greatly consistent with the non-modulated simulations' power spectrum, and hence the results given in 
Table~\ref{tab:all-mm-joint-tests} do not result from the underlying power spectrum discrepancies. 

We are unable to reject the possibility of modulation with such amplitude applied only to the large scales
($\ell \leq 40$). Such modulation is therefore consistent with the GRF field or unnoticed by the 
test (Table~\ref{tab:all-mm-joint-tests}).
However, according to the best-fit $\Lambda$CDM model (assuming even a noiseless observation), 
the multipoles $\ell \leq 40$ carry only about $24\%$ of the map's power.
The possible modulation signals at these scales must also be more difficult to constrain as these are dominated by the cosmic 
variance uncertainty.

To investigate this further, we test simulations with only the large scales being modulated according to $A_{40}=0.114$ 
along direction \lb{225}{-27}. We filter out these scales up to $\lmax{=40}$ (using the \kp sky cut) 
in SH space and downgrade the map to $n_s=64$, 
and process these using a new set of \mm{s} of type: {\it LB} 1 2 - i.e.
having only two regions, each covering a hemisphere. We use 96 such \mm{s} with orientations defined by the centres
of pixels of the northern hemisphere in the ring ordering of the Healpix pixelization scheme of resolution $n_s=4$.
We prepare a set of $1\,000$ modulated simulations treated as data and use $1\,000$ independent
GRF simulations to test the consistency with SI. We split the GRF simulations into two sets of $500$ simulations each, to derive the
covariance matrix, and probe the underlying $\chisq$ distribution.\footnote{Note that with the {\it LB} 1 2 \mm{s} having only two regions it is not 
necessary to process a very large number of simulations to assess a good convergence.} We carry out the multi-region analysis using 96 \mm{s} {\it LB} 1 2, 
and record the values of minimal probabilities (per modulated simulation) and the corresponding orientation of the \mm{}.
The spatial distribution of these orientations defines the accuracy  to reconstruct the correct intrinsic modulation field orientation 
at these scales, under cut sky and negligible amounts (at these scales) of noise.

\begin{figure}[!t]
\centering
\renewcommand{\figurename}{Fig}
\includegraphics[angle=0,width=0.239\textwidth]{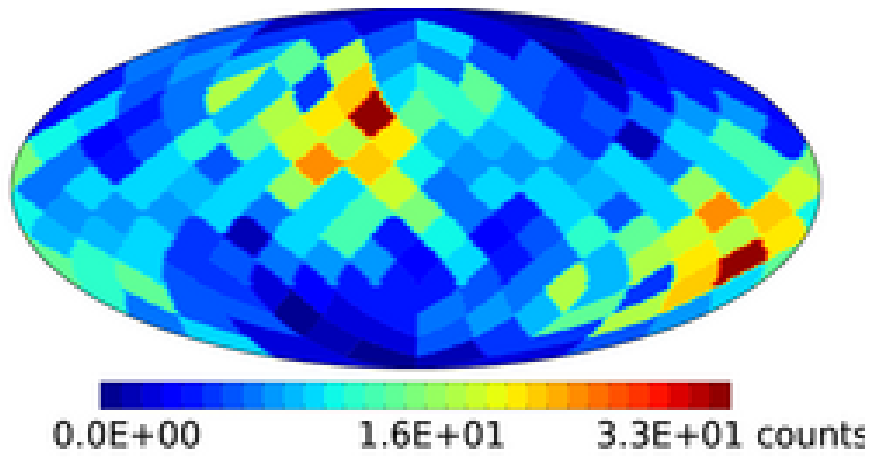}
\includegraphics[angle=0,width=0.239\textwidth]{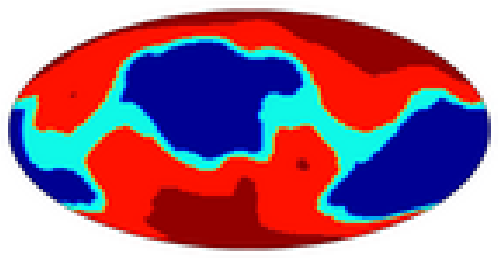}\\
\includegraphics[angle=0,width=0.45\textwidth]{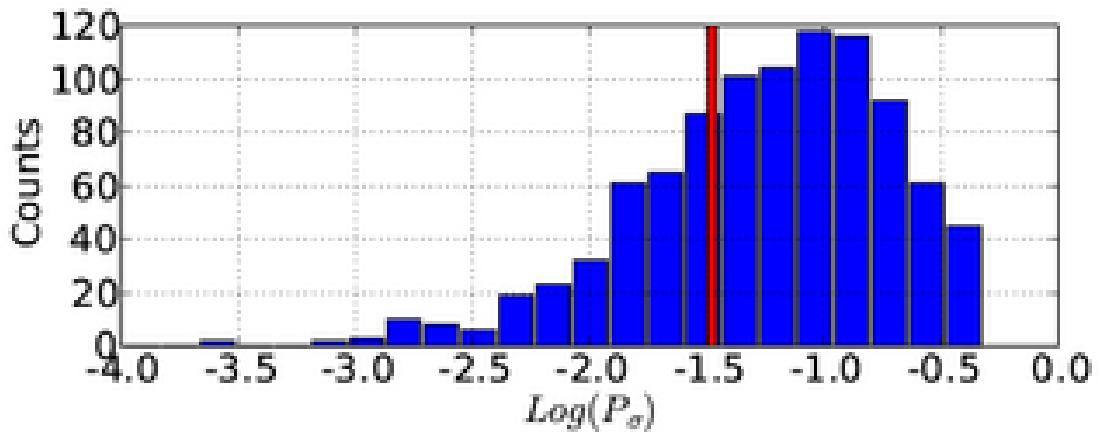}
\caption{Histogram of the reconstructed modulation axis orientations as measured via the  minimal joint
probability $P\bigl(\chisqq\bigr)$ (Eq.~\ref{eq:Pjoint}) in the set of 96 {\it LB} 1 2 \mm{s} from 1000 modulated simulations (top-left) visualized using Healpix grid of resolution $n_s=4$.
The corresponding, reconstructed $50\%$ (dark blue), $68\%$  (light blue) 
and $95\%$ (light red) confidence level contours obtained after smoothing the histogram with a Gaussian beam of FWHM$=7^\circ$ 
(top-right). The reconstructed distribution is normalized on a hemisphere.
In the bottom plot: the distribution of minimal log-probabilities ($\log P\bigl(\chisqq\bigr)$) obtained from 1000 simulations modulated with amplitude
$A_{40}=0.114$ along with the value obtained from the \incv{}  map of the WMAP3 data.
}
\label{fig:mod_axis_orientation}
\end{figure}

The result is plotted in Fig.~\ref{fig:mod_axis_orientation} (top-left). It is easily seen that while the direction is quite correctly
found, the dispersion of the directions within even $50\%$ CL contour (Fig.~\ref{fig:mod_axis_orientation} top-right) is quite large which precludes a very precise determination of the modulation axis.
We find, that statistically $\sim 8 \%$ (the probability corresponding to the peak value in the bottom plot in 
Fig.~\ref{fig:mod_axis_orientation}) of Gaussian simulations, 
to which we compare the modulated simulations treated as a data,
exhibit a more unusual configurations of hemispherical variance distribution.

Consequently, we report this median rejection threshold of $\sim 92\%$, as the statistical sensitivity of the method for detecting
the large scale modulation (i.e. in the filtered maps modulated with $A_{40}=0.114$).
We consider this result - i.e. the low rejection confidence level - to be penalized mostly by the cosmic variance
uncertainty and freedom of phases to assume an arbitrary orientations with respect to the (unsymmetrical) sky cut.
Consequently, we note that due to these uncertainties, it may be difficult to increase this rejection level 
for scales of $\lmax{\sim 40}$ and amplitude $A=0.114$.

\section{Application to WMAP three-year data}
\label{sec:results}

We now present the results of the statistics described in Sect.~\ref{sec:statistics} applied to the three-year and five-year WMAP data.

\subsection{Individual region statistics}
\label{sec:results-single}

\begin{figure}[!t]
\centering
\renewcommand{\figurename}{Fig}
\begin{tabular}{ll}
{\small mean} & {\small variance}\\
\includegraphics[width=0.239\textwidth]{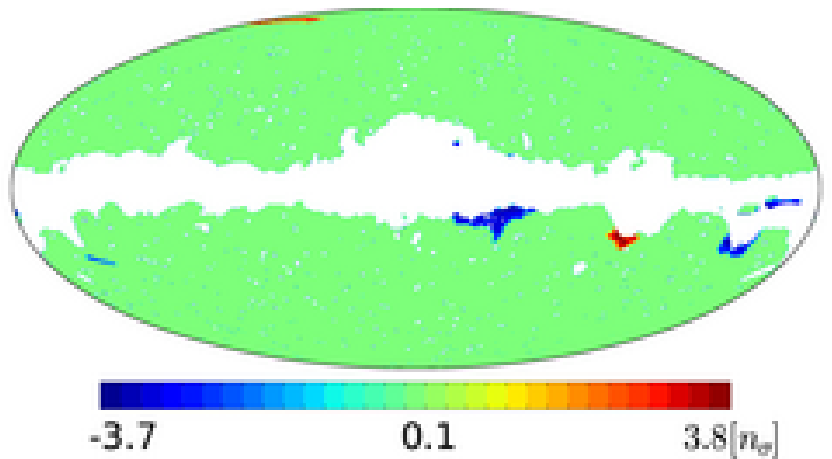}&
\includegraphics[width=0.239\textwidth]{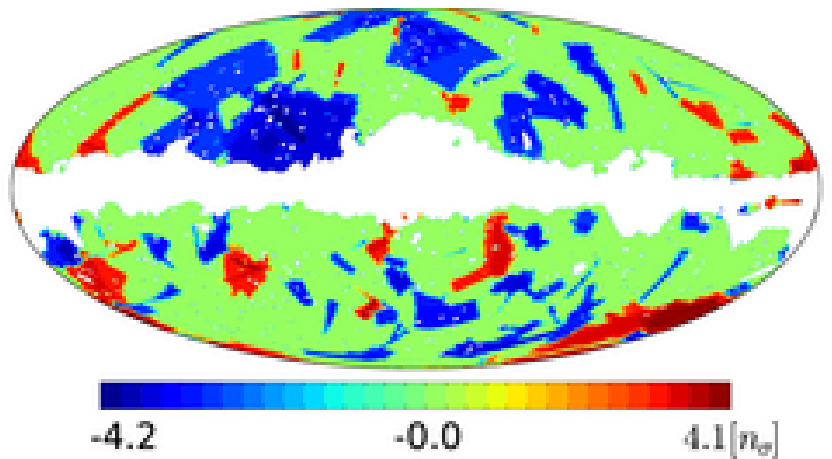}\\
{\small skewness} & {\small kurtosis}\\
\includegraphics[width=0.239\textwidth]{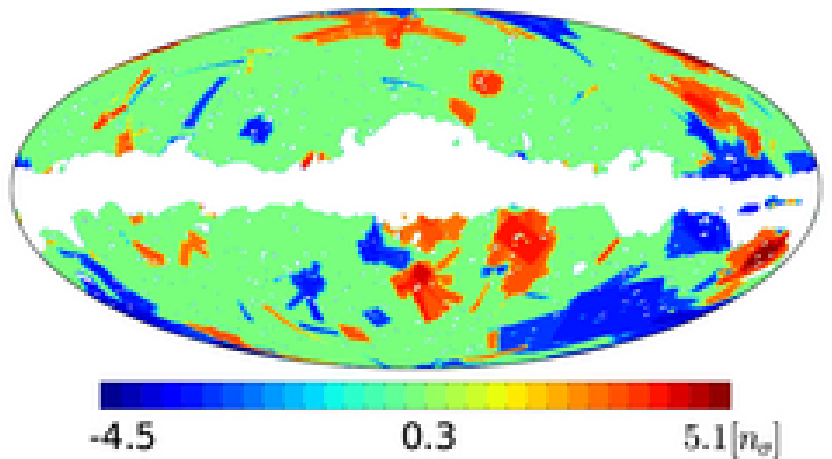}&
\includegraphics[width=0.239\textwidth]{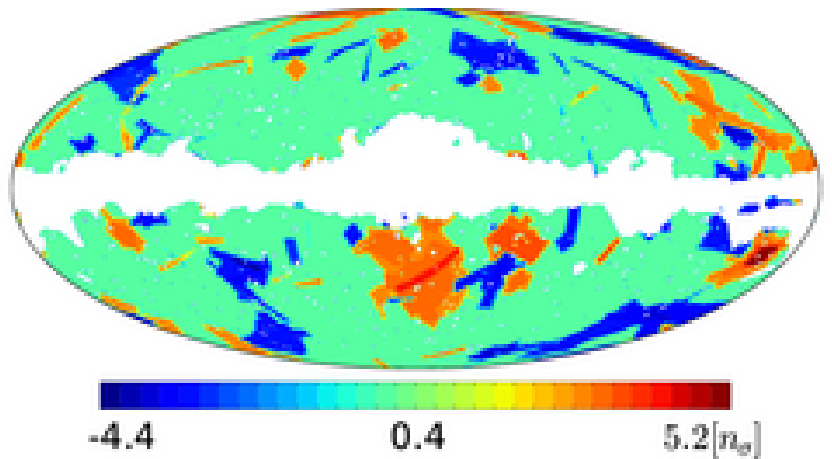}\\
\end{tabular}
\caption{Results of the single-region analysis visualized in the 
  composite $\nsig{ }$ maps of the \incall{ } data, for each of the MODs.
The threshold of $\nsigth{}=3$ (as defined in Sect.~\ref{sec:visualizing_results}) is used.
Mollweide projection and the galactic coordinates are
  used. The origin of the coordinate system $(l,b)=(0,0)$ is in center
  of the plots and the galactic longitude increases leftwards. Regions
  around the Galactic plane are partially removed by the \kp sky
  mask. The same convention is kept throughout the rest of this
  paper. 
}
\label{fig:all-single-reg}
\end{figure}

The  individual region statistics as described in
Sect.~\ref{sec:statistics} find numerous regions amongst our many
\regsk{s}, which deviate by more than certain $\nsigth{|}$ in all MODs. Table~\ref{tab:single-reg-3sdet} gives an
incomplete list of some of the strongest detections. In
Fig.~\ref{fig:all-single-reg} we plot the detected regions at
$\nsigth{|} = 3$ in the individual region statistics of the \incall{} data.
\begin{figure}[!t]
\centering
\renewcommand{\figurename}{Fig}
\includegraphics[width=0.5\textwidth]{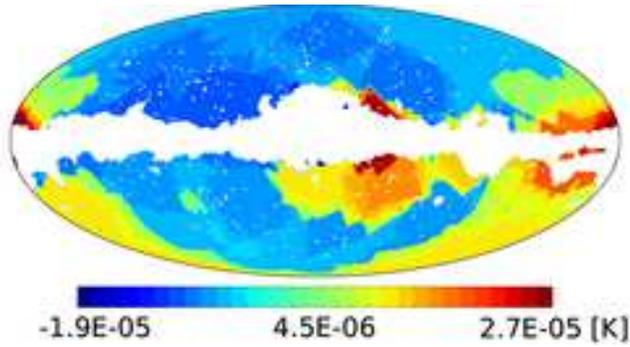}
\caption{Results of the single region analysis.
Residual difference ($\Delta=\sqrt{\sigma^2}-\langle\sqrt{\sigma^2}\rangle$) scrambled map of the variance distribution in the \incall{} map processed with
100 \mm{s} of the {\it HP} 2 \regsk{}. The well know, large scale
hemispherical power distribution asymmetry is clearly seen as
  is the distribution of foreground residuals along the galactic cut.}
\label{fig:all_hp2-single-diff-s}
\end{figure}
\begin{table}[!b]
\caption{An incomplete list of strongest deviation directions from maps in Fig~\ref{fig:all-single-reg} at $\nsigth{|} =3$ in the individual 
region analysis for means, variances, S and K, sorted in galactic
longitude ascending order. Notice that with our simulations we
directly probe the $\pm 3.7\sigma$ PDF region without need for
extrapolations.}
\centering
\begin{tabular}{llllllll}\hline\hline
\multicolumn{2}{c}{mean}&\multicolumn{2}{c}{variance}&\multicolumn{2}{c}{skewness} &  \multicolumn{2}{c}{kurtosis}\\
(l,b) & $\nsig{}$ & (l,b) & $\nsig{}$ & (l,b) & $\nsig{}$ &(l,b) & $\nsig{}$ \\\hline
157, -29 & -3.07           & 63, 28 & -4.21  & 173, -73 & -4.50 & 84, -31 & -4.38 \\
\bf{211, -57} & \bf{-2.78} & 67, 19 & -4.21  & 193, -26 & 5.06  & 195, -27 & 5.20 \\
241, 42 & -2.36            & 199, -55 & 4.13 & 209, 8 & -3.50   & \bf{212,-55} & \bf{-3.77} \\
265, -21 & 3.81            & 319, -27 & 3.40 & 217, 35 & 4.00   & 309, 59 & -3.25 \\
318, -9 & -3.69            &          &      & 225, -20 & -3.60 & 312, -21 & 3.65 \\
167, 79 & 3.06             &          &      & 241, -52 & -3.51 & 356, -36 & 4.25 \\
        &                  &          &      & 311, -20 & 3.86  &          &      \\
        &                  &          &      & 357, -35 & 4.15  &          &      \\\hline
\end{tabular}
\label{tab:single-reg-3sdet}
\end{table}

Note that with our approach the so-called ``cold spot'' is not
detected at $3\sigma$ level from being cold (i.e. via distribution of means) at all tested resolutions (Table~\ref{tab:single-reg-3sdet}). It appears at about $\sim
2.7\sigma$ around galactic coordinates \lb{211}{-57}.  Excessively
``cold'' or ``hot'' deviations in all MODs are detected in general
with large values of $\nsig{}$. 

As expected the $\nsig{}$ map of the means in Fig.~\ref{fig:all-single-reg} shows that the strongest deviating
regions are directly close to the galactic plane cut off by the \kp
mask, thus hinting at foreground residuals.

The variance $\nsig{ }$ map shows local strong anomalies with the extended variance suppression
in the northern hemisphere towards \lb{67}{19} and with an extended variance excess towards \lb{199}{-55}. 
We note that these localized anomalies must, at least in a part, make up for the hemispherical power asymmetry.

Skewness and kurtosis maps consistently
indicate strong local deviations from GRF simulations towards \lb{193}{-26} and
\lb{356}{-36}. While some regions appear in all three maps, some
appear only in one of the moments therefore the correlation between
those results is not obvious.  

\begin{figure}[t!]
\centering
\renewcommand{\figurename}{Fig}
\includegraphics[angle=-90,width=0.239\textwidth]{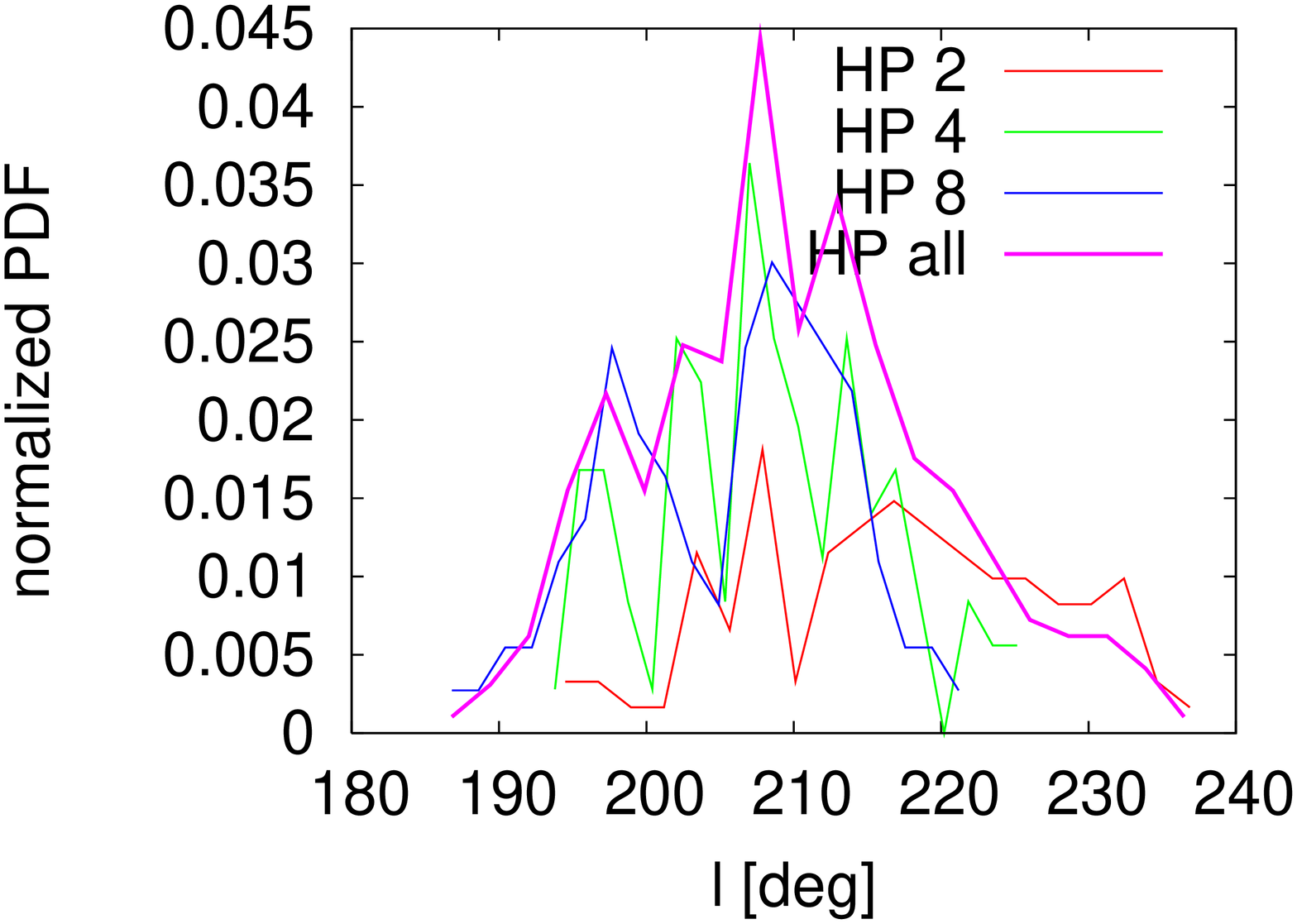}
\includegraphics[angle=-90,width=0.239\textwidth]{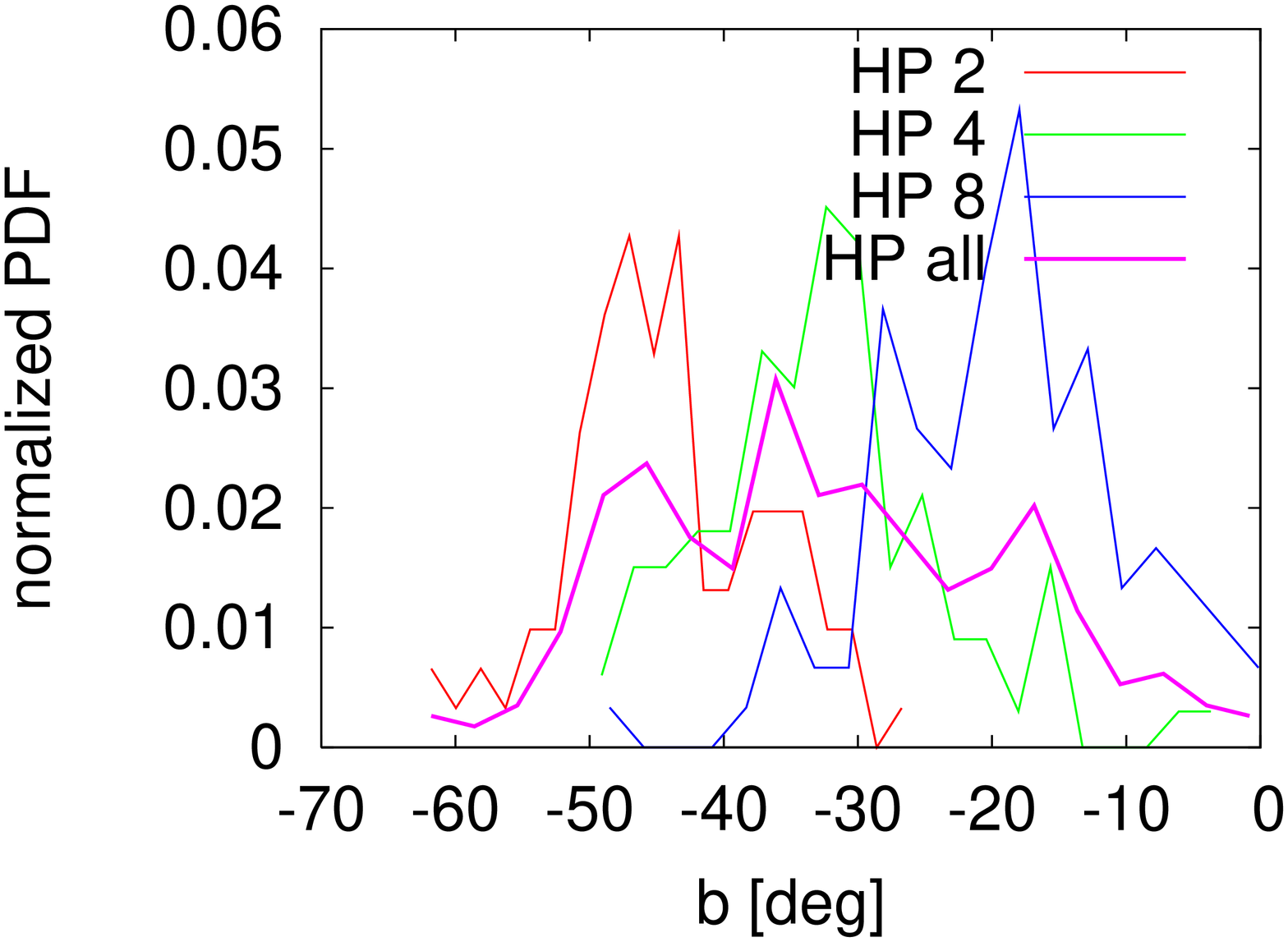}
\caption{Orientation of the dipole component (galactic longitude -
  left panel and galactic latitude - right panel) of the $\nsig{}$
  maps for the \incall{} map processed individually with all {\it HP} \regsk{s}. Each color corresponds to a different resolution of the \regsk{}.  
While the longitudinal orientation of the dipole does not vary with
resolution, the galactic latitude systematically shifts to lower galactic latitudes as resolution increases.}
\label{fig:all_hp2_dipole_PDFs}
\end{figure}

In order to investigate the spatial distribution of this high $\nsig{}$
regions,  we plot in Fig.~\ref{fig:all_hp2-single-diff-s} the $\Delta$ map, of differences
between the variance distribution measured in each region of  \mm{s} of
the {\it HP} 2 \regsk{}, and the simulations average,
$\Delta\equiv d_i=\sqrt{\sigma_i^2}-\langle\sqrt{\sigma_i^2}\rangle$. The map is 
obtained by scrambling 100 difference maps from 100 different \mm{s} (as described in Sect.~\ref{sec:visualizing_results}). 
This map can be seen as a residual map for the local variance. This residual
map exhibits a well known power asymmetry
\citep{2004ApJ...605...14E,2007ApJ...660L..81E}. The dipole ($\ell=1$)
component  of this residual masked map (ignoring the effect of the
mask) is aligned along axis \lb{237}{-44} with power excess in the
southern hemisphere. In order to probe this direction  further, for each individual \mm{} we also produced an 
$\nsig{}$ estimator maps and checked the orientation of the dipole
axis. Fig.~\ref{fig:all_hp2_dipole_PDFs} shows the  PDFs obtained for
the orientation of the dipole asymmetry in galactic longitude and
latitude. Interestingly, we notice that the orientation of the axis of the 
hemispherical power asymmetry has some scale dependence.
When using smaller scales with finer \regsk{s}, the orientation of the power asymmetry dipole shifts
from larger galactic latitudes (roughly from the position of the cold
spot, with the mean PDF value ~\lb{218}{-43}   -- see also
Table~\ref{tab:single-reg-3sdet}) for the {\it HP} 2 resolution to smaller latitudes
~\lb{206}{-18} for the {\it HP} 8 \regsk{}. 
The dependence of the power asymmetry orientation 
in function of the pre-filtered in SH space data have previously been tested by \cite{2004MNRAS.354..641H}.
While we will return to the
power asymmetry issue in the next subsections, we note that the medians
of the dipole axis distributions of other MODs  are not correlated
with the dipole axis orientation of the variance map
(Fig.~\ref{fig:all_hp2_dipole_PDFs}) and generally point at some other locations. 

In the next section we quantify the statistical significance of these deviations.

\subsection{Joint multi-region statistics}
\label{sec:results-multi-region}

\begin{figure*}[!t]
\centering
\renewcommand{\figurename}{Fig}
\includegraphics[width=\textwidth]{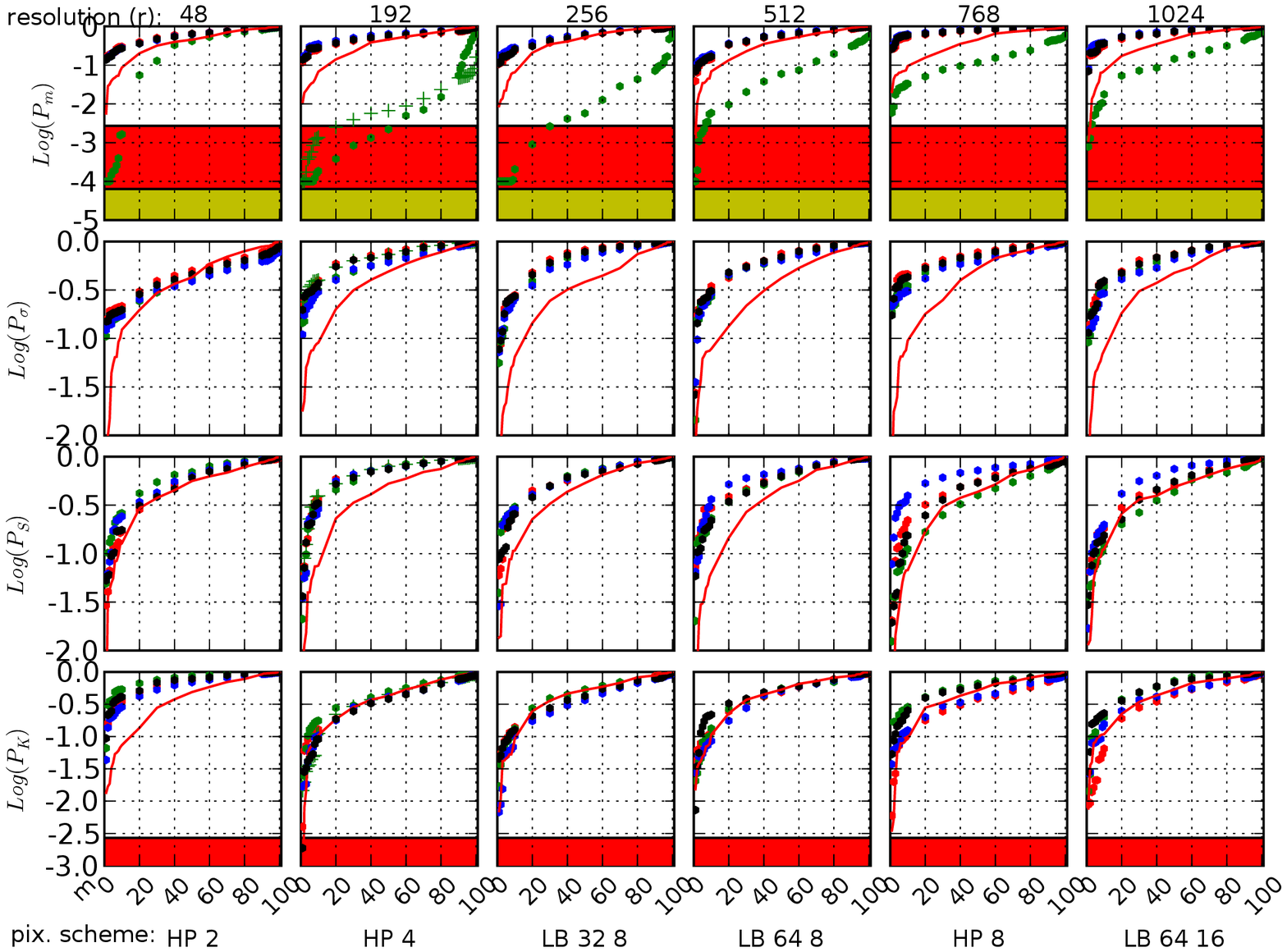}
\caption{ Results of the ``multi-region'' analysis.
Dependence of joint probabilities of exceeding ($P\bigl(\chisqq\bigr)$, Eq.~\ref{eq:Pjoint}) as a function of \mm{} number
for all \regsk{s} considered.
The probabilities calculated with the spectral data maps from channels
Q, V, W, \incall{ } are plotted in red, green, blue and black dots respectively.  
Each dot corresponds to the joint probability using one \mm. From the left to
the right, the panels show results for increasing resolution of the \regsk{} (see Table~\ref{tab:regionalizations}) 
with 100 different \mm{s } (along abscissa) in each. 
From top to bottom the four rows correspond to the four MODs - i.e.
mean, variance, S and K respectively. 
Probabilities corresponding to the WMAP5 \incv{5} data for \regsk{} {\it HP} 4 are plotted with green crosses ($+$).
$3\sigma$ and $4\sigma$ confidence levels are shaded in red and yellow respectively. 
The joint probabilities were sorted in each dataset before plotting for better visualization;
therefore the probabilities from different datasets generally do not correspond to the same
\mm{} numbers and do not directly correspond to the unique reference
numbers used in the analysis. Hence the most unlikely events are localized at the left side in each panel.
Additionally we plot the thin red line, which indicates the distribution of probabilities obtained from 100 GRF simulations
of the \incq{} data, each of which was processed with one (different for each simulation) \mm{}. If the data
follow the expectation of GRF then statistical departure of data from this line would manifest certain degree of correlation
between probabilities obtained with different \mm{s} for the same dataset (as discussed in Sect. \ref{sec:discussion}).
For better visualization in range $m\in[10,90]$ we plot only every 10'th sorted probability value.
}
\label{fig:all-results}
\end{figure*}

In Fig.~\ref{fig:all-results} we plot a compilation of all joint
``probabilities of exceeding'', calculated with all datasets considered (\incq, \incv,
\incw, and \incall)  in all 600 \mm{s}. In order to visualize the smallest probabilities
logarithmic scale is used. 
Note that we sort these probabilities in each MOD
and dataset so as to ease visualization, so the points with same
abscissa in different MOD and data sets do not necessarily correspond to the same \mm{}. 
  
Most of the results concentrate along the zero point of the joint log-probabilities, which indicates a
good consistency of the data with the simulations at relatively high
CL. (The white region in the Fig.~\ref{fig:all-results} encompasses
CLs of up to $3\sigma$;  $3\sigma$ and $4\sigma$ regions are shaded in
red and yellow respectively). 

It is important to note that within one \regsk{} the dispersion 
of probabilities in the Fig.~\ref{fig:all-results} results only from the orientation of the \mm{}.
As a result, the statistical method involving many \regsk{s} help us obtain the unbiased results that one could get
relying only using a single \regsk{}.
We also recall that for each plotted point, the statistic was also calculated for $\NsimMCPDFq=1\,000$ simulations 
in order to probe the underlying PDFs. For each point the corresponding full covariance matrix was obtained from
$\NsimCq=9\,000$ simulations (as described in Appendix~\ref{sec:multi_region}).\\

We now focus on three distinctive sets of results, based on
the Fig.~\ref{fig:all-results}, and quantify the deviations in more
detail. We detail on a tentative excess seen in kurtosis before focusing
on the large scale power anomalies, and on unusually strong dipole
contribution in the V channel of the WMAP. 
Then, we comment on the results from tests carried out with the difference map datasets.

\subsubsection{Localized Kurtosis excess}
\label{sec:NGK}
In Fig.~\ref{fig:all-results} there is one ``$3\sigma$'' detection in kurtosis in {\it HP} 4 \regsk{} in the \incall{} dataset 
(bottom second from the left panel in Fig.~\ref{fig:all-results}) --
a result found using one in $100$ of \mm{s} probing these scales.
Here we discuss this particular point as a tentative detection because
although the \mm{} bins the data to create the most unlikely realization of the kurtosis, 
it lies in the low-end tail of the whole spectrum of equivalent measurements and hence its statistical impact cannot be large.

\begin{figure}[!t]
\centering
\renewcommand{\figurename}{Fig}
\begin{tabular}{l}
a)\\ \includegraphics[width=0.45\textwidth]{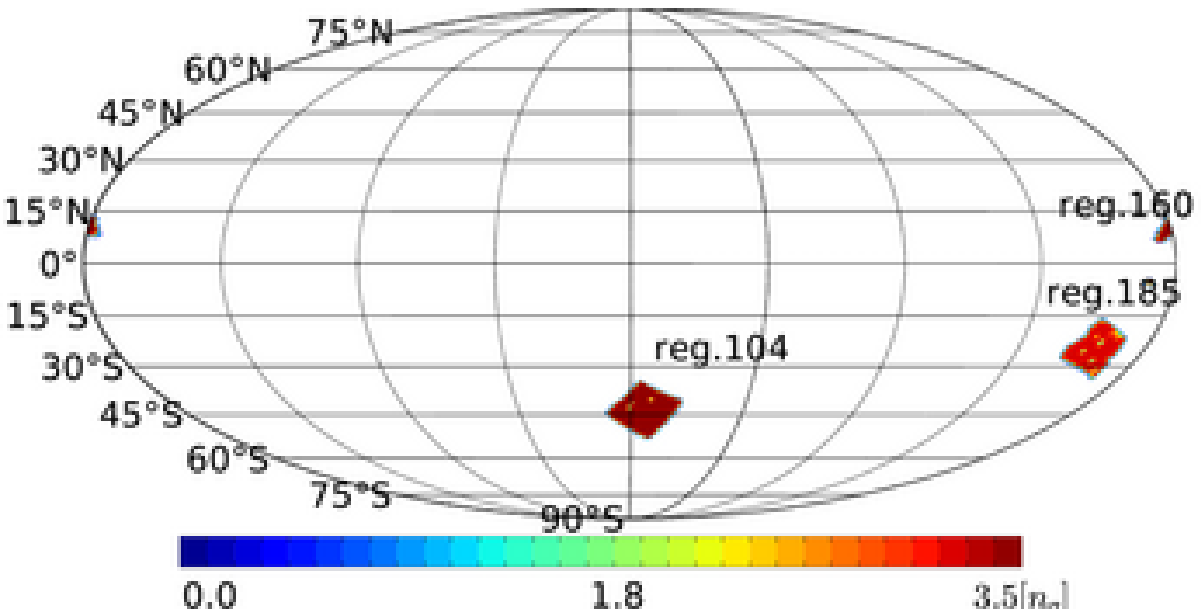}\\
b)\\ \includegraphics[width=0.45\textwidth]{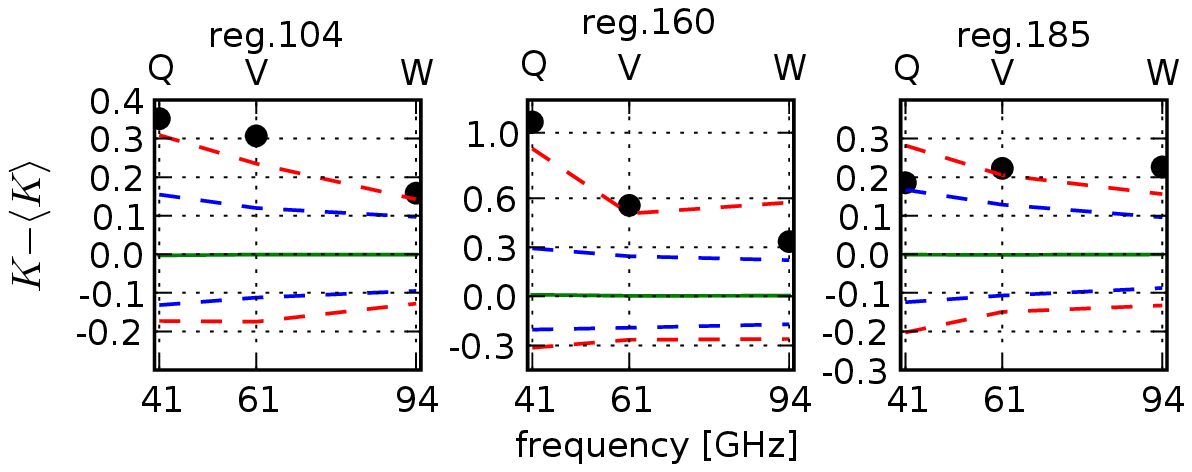}
\end{tabular}
\caption{a) Kurtosis (thresholded at $3\sigma$) $\nsig{}$ estimator map from the multi-region
analysis of the \incall{} data, inconsistent with GRF hypothesis at joint probability
$>99.73\%$ CL. Only one multimask (\mmID{53} of the {\it HP} 4 \regsk{}) is used for this map, since only the one out of $\Nr\Nmm=600$ 
yields $|P(\chisqq)|<\Pth=1-0.9973$. (See text and Table~\ref{tab:NGdet-info} for more details).
b) The spectral dependence of the kurtosis in the depicted regions in the three WMAP frequency bands along with the 
two and three sigma contours, from 1000 simulations, and the simulations mean are also plotted.
}
\label{fig:all-multi-reg-detNG}
\end{figure}
\begin{table}[bth]
\caption{Three $\sigma$ NG detections in K multi-region statistics of the \incall{} data 
using \mm{} resulting in joint probability $P\bigl(\chisqq\bigr)<0.0027$ for the resolution $n_s=512$ and
$P\bigl(\chisqq\bigr)<0.005$ for the resolution $n_s=64$.
The (l,b) field gives the galactic coordinates to the center of the region
}
\centering
\begin{footnotesize}
\begin{tabular}{ccccccc}\hline\hline
&\multicolumn{2}{c}{\makebox[0.15\textwidth][l]{resolution ($n_s=512$)}} & \multicolumn{2}{c}{\makebox[0.15\textwidth][l]{resolution ($n_s=64$)}}\\
\makebox[0.05\textwidth][c]{region}& \makebox[0.07\textwidth][c]{$\Npix$}  & $\nsig{}$ & \makebox[0.07\textwidth][c]{$\Npix$}  & $\nsig{}$ & (l,b)\\\hline
160 & 2750 & 3.48   &  37  & 1.38 & 181 , 2\\
185 & 12610 & 3.21  & 199  & 3.26 & 199 , -23\\
104 & 16125 & 3.52  &250  & 3.27 & 355 , -44
\end{tabular}
\end{footnotesize}
\label{tab:NGdet-info}
\end{table}
In Fig.~\ref{fig:all-multi-reg-detNG}a we plot, thresholded at ``$3\sigma$'', the $\nsig{}$ map using only this particular \mm{}. 
The details of the three most strongly deviating regions in this map are given in
Table~\ref{tab:NGdet-info}. 
In case of two other \mm{s}, using which the \incall{} data yield a detection at confidence levels of 99\% and 98\% in K 
({\it LB} 64 8) and S ({\it HP} 8) respectively,
the ``$3\sigma$''-deviating regions turn out to be similarly located (like those in Fig.~\ref{fig:all-multi-reg-detNG}a).
With these regions masked out from the analysis a good consistency with the GRF simulations is reached. 
Note that the region 160 is rather small -- mostly removed by the galactic sky cut 
(see Fig.~\ref{fig:all-multi-reg-detNG}a and Table.~\ref{tab:NGdet-info} for precise coordinates and size) 
however masking it has a comparable effect on the joint
probability increase as masking out the two other and much larger regions. 
The consistency of the \incall{} data with GRF 
simulations increases from 0.18\% (without removal) to ~1.1\%  and 0.9\% with regions 160 and 104+185
removed respectively.  
Individual removal of regions 185 and 104 only increases the level  of consistency by $\lesssim 0.5\%$. 
The simultaneous removal of all three regions increases the consistency up to 12\%. 

\paragraph{Dependence as a function of \mm{}}

To further test the robustness of this detection
we have generated two other {\it HP} 4 \regsk{} sets of $\Nmm=100$ \mm{} each: one 
by simply choosing the three rotation angles with the prescriptions given earlier, 
and the other by focusing only on the region in the rotation angle parameter space within $\pm 5^\circ$  
around the original orientation of the \mm{} leading to the $3\sigma$ detection.

With the first set we obtained results yielding a joint probability $P(\chisqq)<0.05$ with 3 \mm{s}, while in the second
we find that 25\% of \mm{} yield $P(\chisqq)<0.05$, and 4\% yield $P(\chisqq)<0.01$, with the
  strongest detection $P(\chisqq)=0.0035$, of which the $\nsig{}$ map
  points to the same three regions as depicted in Fig.~\ref{fig:all-multi-reg-detNG}a. 
We note that the reported regions (160 and 185) are located in directions towards which the strongest deviations in the individual 
region  statistics  (Fig.~\ref{fig:all-single-reg}) were found.

\paragraph{Dependence as a function of frequency and resolution }

In Fig.~\ref{fig:all-multi-reg-detNG}b we present the
spectral dependence of kurtosis in the regions depicted in Fig.~\ref{fig:all-multi-reg-detNG}a.
While there is a non-trivial spectral dependence in regions 160 and 185, with opposite tilt -- red and
blue respectively --
there is almost no spectral dependence in region 104.

We also check the dependence on the S/N ratio in the selected regions.
For this purpose we downgrade all datasets and simulations to
resolution $n_s=64$, which effectively increases the S/N ratio per pixel by
a factor of 8. We redo the multi-region analysis lowering the minimal region pixel number threshold down to $N_{pix}>10$ 
and find that the minimal probability per \mm{} ($P\bigl(\chisqq\bigr)$)
corresponds to a rejection threshold of $99.5\%$ CL (Table~\ref{tab:NGdet-info}).  
As seen in Table~\ref{tab:NGdet-info} the individual region response to the resolution change
is strong only in case of region 160 while in the two other regions it is rather small. 
The result is robust under variations of region pixel number threshold
and the number of simulations used to probe the underlying PDFs ($\NsimMCPDFq\in\{1000,5000\}$). 
Masking-out regions 104 and 185 reduces the anomaly to
$\sim 96\%$ CL and as expected in this case, removal of region 160 has basically no impact on this value.

\paragraph{Summary}
The non-trivial spectral dependence and close galactic-plane alignment in two of the three selected regions (160 and 185)
suggests presence of some residual foreground anomalies.
In case of the regions away from the galactic plane (104 and 185)
since the local oddity is insensitive to the S/N ratio change 
it is also unlikely that an unknown instrumental noise fluke generates them.
While we will return to the overall statistical significance of these findings in Sect.~\ref{sec:jointall}, 
we note that the positive KE seems to be inconsistent with the
extended foregrounds interpretation of these detections, according to the results from Sect.~\ref{sec:tests}.
Also the $3\sigma$
detection of the multi-region analysis appears only in the \incall{} data, but is clearly weaker in other single band maps.

Given that this detection results from just one particular \mm{} and is selected from the lower-tail end of a whole
distribution of equivalent measurements, it is inconclusive as regards indicating whether this detection is not just a fluke. 
Given that, we report in particular region 160, whose removal leads  the overall significance to drop below $3\sigma$ CL,
as a tentative detection noting that more sophisticated local statistical analyses (see Sect.~\ref{sec:discussion}) 
could be invoked to back these results up or refute them.
\begin{figure*}[!t]
\centering
\renewcommand{\figurename}{Fig}
\includegraphics[width=\textwidth]{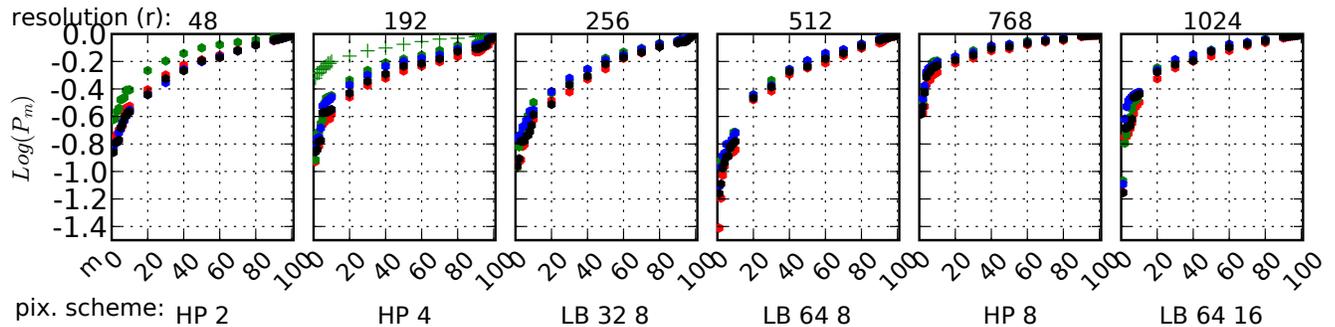}
\caption{Results of the ``multi-region'' analysis with the dipole-corrected \incv{} (\incv{5}) data.
Joint probabilities $P\bigl(\chisqq\bigr)$ (Eq.~\ref{eq:Pjoint}) for 600 \mm{s}. 
The probabilities calculated with the spectral data maps from channels
Q, V, W, \incall{ } are plotted with red, green, blue and black dots respectively.  
WMAP5 \incv{5} data are plotted with green crosses (+) in {\it HP} 4 only.
The V dataset original dipole has been replaced  by the simulated
dipole and removed in case of \incv{5} data. 
Each dot (cross) corresponds to the joint probability of one \mm. From left to the right, the
panels show the results with increasing resolution of \regsk{s} (see
Table~\ref{tab:regionalizations}) with 100 different \mm{s } in each
panel.  Only the ``mean'' data is shown since all other results remain almost unchanged. 
For better visualization in range $m\in[10,90]$ we plot only every 10'th sorted probability value.
}
\label{fig:all-results-vl0..1}
\end{figure*}

\subsubsection{Variance large scale distribution}
\label{sec:SGsigma}
\par In Sect.~\ref{sec:results-single} we analyzed the large
scale power distribution as measured via \mm{s} with regions of angular sizes ranging from $6^\circ$ to $30^\circ$
(Fig.~\ref{fig:all_hp2-single-diff-s} and Fig.~\ref{fig:all_hp2_dipole_PDFs}).

In this section we focus on the joint multi-region analysis  
of the variance distribution. The corresponding results are illustrated in the second row
of Fig.~\ref{fig:all-results}. The data remain in excellent consistency with the simulations.

In Sect.~\ref{sec:sensitivity_to_power_anomalies}  we found that the modulation amplitude (defined by Eq.~\ref{eq:power_mod}) of $\Almax{=1024}=0.114$ 
would be rejected at $>99.9\%$ CL, and we argued that the modulation parameter  $\Almax{=40}=0.114$ would statistically be difficult to exclude at CL higher than
$92\%$. 
Using our main set of the \mm{s} (Table~\ref{tab:regionalizations}) 
we fail to detect, in any of the data sets tested, any statistically significant anomaly, such as the claimed hemispherical power asymmetry 
(depicted in Fig.~\ref{fig:all_hp2-single-diff-s}), as measured from the large scale variance distributions in the multi-region analysis.

\begin{figure}[!t]
\centering
\renewcommand{\figurename}{Fig}
\begin{tabular}{cc}
{\small WMAP3 \incv{} } & {\small WMAP5 \incv{}}\\
\includegraphics[width=0.239\textwidth]{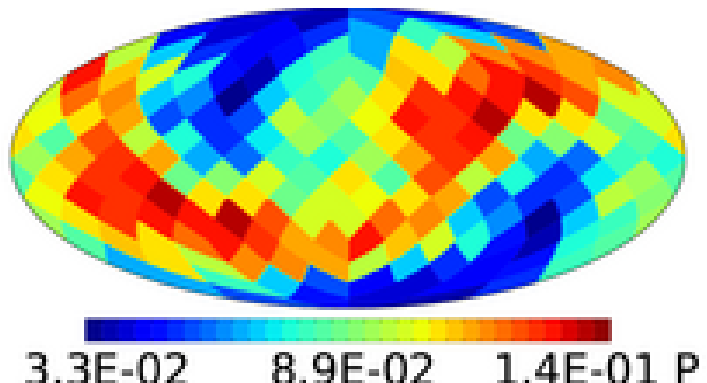}&
\includegraphics[width=0.239\textwidth]{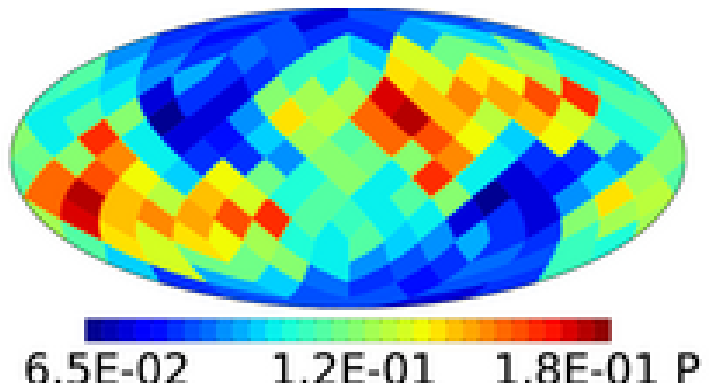}\\
\end{tabular}
\caption{Results of the multi-region analysis, using the set of 96 {\it LB} 1 2 \mm{s}, of the filtered in SH space up to $\lmax{=40}$, 
low-resolution \incv{} data (left), and \incv{5} data (right). 
The color in each pixel encode the multi-region, joint ``probability of exceeding'' derived with
the {\it LB} 1 2 \mm{} rotated to the direction of the center of that pixel. Note that half of pixels in each map is redundant.
}
\label{fig:wmap3wmap5-LB_1_2}
\end{figure}

As an extension to that, we repeat the analysis performed in Sect.~\ref{sec:sensitivity_to_power_anomalies} 
for the low resolution, filtered in SH space up to $\lmax{=40}$, 
WMAP data, using the same set of {\it LB} 1 2 \mm{s},
to test the variance distributions in the corresponding set of 96 differently oriented hemisphere pairs.
We thereby extend the test for the largest possible scales of $180^\circ$.
We merge the \kp sky mask with the {\it LB} \mm{s} for the analysis of the WMAP3 data, and KQ75 sky mask for analysis of the WMAP5 data.
The result of the multi-region (here only two region) analysis is plotted in Fig.~\ref{fig:wmap3wmap5-LB_1_2} 
for the \incv{} data (left) and \incv{5} data (right).
The minimal ``probabilities of exceeding'' found are: $\min{(P(\chisqq{}))}\approx 3.3\%$ (also marked in Fig.~\ref{fig:mod_axis_orientation} bottom)
towards \lb{247.5}{-30}\footnote{This and the following result is accurate to within the tolerance of about $\sim \pm 7^\circ$ resulting from the 
low-resolution search involving only 96 directions over a hemisphere in the {\it HP} 4 \regsk{}.}
for the \incv{} data and $\min{(P(\chisqq{}))}\approx 6.5\%$ towards \lb{281.5}{-19} for the \incv{5} data.

These two results agree well with the previously estimated \citep{2007ApJ...660L..81E} intrinsic modulation parameter value $A=0.114$ at
scales $\lmax{\leq 40}$, as they lie well within ``one-sigma'' region of the distribution of log-likelihoods obtained 
from 1000 simulations modulated with the modulation of $A_{40}=0.114$. However we note that it will always be difficult to reject
such modulation at high confidence level as it is also realized (to this or a greater extent) 
on average in $\sim 8\%$ of GRF simulations (Sec.~\ref{sec:sensitivity_to_power_anomalies}).

Note that the distribution of the probabilities of the joint multi-region analysis (Fig.~\ref{fig:wmap3wmap5-LB_1_2})  
has a very flat and extended maximum and, for example,  the minimal joint-probability in the \incv{} data in the reported direction
is only $0.2\%$ smaller from the probability corresponding to the direction close to the galactic pole, which is roughly $50^\circ$ 
away from the minimal probability direction.

Analogous analysis, involving the {\it LB} 1 2 \mm{s}, but performed on the full resolution unfiltered WMAP \incv{5} data,
results in larger minimal-probabilities: $\min{(P(\chisqq{}))}\approx 9.6\%$ and the probability is minimized towards \lb{225}{-78}.

\subsubsection{Residual dipole of the WMAP V channel.}
\label{sec:NGV}

In Fig.~\ref{fig:all-results} we see a deviation in the distributions
of the mean in the \incv{} dataset (green dots) and in \incv{5} dataset (green crosses) for  most of the
\mm{s}. The fact that it is visible in most  \mm{s } suggests that the
anomaly is not particularly sensitive to the \mm{} orientation, and
that it comes from large angular scales. Indeed, as measured by the 
$\nsig{}$ values in individual regions of the \mm{} with the lowest joint
probabilities, no region significantly deviates from simulations. 

\par However, we find that the \incv{} data are fully consistent with
the GRF simulations if we remove the dipole from the data, which is
roughly $\sim 2$ times larger than the one in our
simulations \footnote{During the final stage of this work this sort of anomaly
 was independently reported in \cite{2008ApJ...672L..87E}}. 
Actually the dipole values in the datasets as measured by the
multipoles $l=1$ on the \kp cut sky power spectrum are: 
$47 ({\rm \mu K})^2$, $54 ({\rm \mu K})^2$, $45 ({\rm \mu K})^2$ 
in \incq{}, \incv{}, \incw{} datasets respectively. 
The corresponding values in the WMAP5 data are: $64 ({\rm \mu K})^2$, $54 ({\rm \mu K})^2$ in \incv{5} and \incw{5} maps respectively.
The measurements of dipoles on initially dipole-free maps, using a sky mask, introduce a 
bias due to power leakage from other coupled multipoles, leading to non-zero dipole amplitudes.
When sky-cut-generated dipoles are statistically accounted for, the result would yield:
$20^{+104}_{-26} ({\rm \mu K})^2$, $27^{+104}_{-26} (38^{+104}_{-26}) ({\rm \mu K})^2$, $18^{+104}_{-26} (27^{+104}_{-26}) ({\rm \mu K})^2$ in the WMAP3 (WMAP5)
data at 95\% CL, and hence is consistent with vanishing intrinsic dipole (except for the V band channel).
We note that the noise component generates dipoles with amplitudes of order $C_1 \sim 0.01 ({\rm \mu K})^2$ which is about three 
orders of magnitude less than the leakage effect.
However the 95\% CL effect is not sufficient to explain the strong anomalies detected in the regional tests.

The anomaly is more visible in the difference of dipole amplitudes between different channels.
The difference of $9\pm 3 ({\rm \mu K})^2$ (at 95\% CL) between channels V and W is excluded 
using simulations at $>99.9\%$ CL assuming that it is generated only by the power leakage from the cut sky.
The difference in the WMAP \incv{5} data is even larger: $11\pm3 ({\rm \mu K})^2$.

\begin{figure}[!t]
\centering
\renewcommand{\figurename}{Fig}
\includegraphics[width=0.45\textwidth]{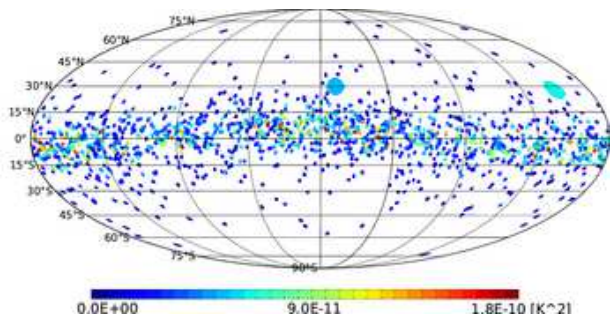}
\caption{ 
Distribution of dipole orientations generated due to \kp cut sky from 1467 full sky simulations with vanishing
initial dipole. The large spot in the center (at the right-hand side) of the plot, 
indicates the orientation of the WMAP3 (WMAP5) V band dipole. The color scale reflects the 
amplitudes of the leakage generated dipoles.}
\label{fig:leakage_generated_dipoles}
\end{figure}

As for the amplitude of the V band dipole again, it becomes anomalous as one considers not only the magnitude, but also its orientation.
The dipoles generated due to power leakage are strongly aligned within the galactic plane (Fig.~\ref{fig:leakage_generated_dipoles})
as a result of the shape of the \kp sky mask.
While the \incq{} and \incw{} dipoles measured on the cut sky are
close from each other and close to the galactic plane pointing at \lb{13}{-8} and \lb{7}{5} respectively,
the dipole of the V band points at \lb{350}{30} which is itself anomalous at $>97\%$ CL.
WMAP5 data yield the dipole orientation \lb{203}{28} (Fig.~\ref{fig:leakage_generated_dipoles}).

We note that all dipoles with $b>30^\circ$ have much smaller (roughly by an order of magnitude) amplitude than the one in the 
V band of the WMAP, which we believe is the reason for strong detections in the regional statistics carried out in the previous sections.
When combining the alignment of the V band dipole with its magnitude, the hypothesis that it's generated only via the power leakage
can be excluded at a very high CL since out of 1300 simulations, and within the subset of 37 that have generated dipole aligned at 
$|b|>30^\circ$ the maximal generated power is of only $13 ({\rm \mu K})^2$ which makes even the CL estimation unfeasible, since this when 
compared to the $54 ({\rm \mu K})^2$  of the \incv{} dataset ($65 ({\rm \mu K})^2$ for \incv{5}), the simple $\chi^2$
test implies a rejection basically without doubt to a very reasonable limits.

By reducing the dipole amplitude to the level consistent with simulations, or
alternatively, by replacing it with our of our simulated dipoles, the data become
consistent with our simulations at $< 2\sigma$ CL in the joint
multi-region analysis (see Fig.~\ref{fig:all-results-vl0..1}) and at
$<1\sigma$ CL in all-\mm{s} analysis at all resolutions  (see
Table~\ref{tab:all-mm-joint} in the next section). Note that the
presence of this dipole in the V band is of no cosmological
consequences since the dipole is marginalized over for any
cosmological analysis, but may be important for other low-$\ell$ analyses.

\subsection{All-\mm{s} analysis}
\label{sec:jointall}
\begin{table}[!t]
\caption{Results of the all-\mm{s} analysis for the signal dominated (co-added) and noise dominated (difference) maps. 
The columns content is as follows: (1) data set, (2) \regsk{}, (3..5) ``probability of rejecting'' the consistency with GRF simulations
(Sect.~\ref{sec:all_joint}) for each MOD.
In case of \incv{} and \incv{5} datasets, the probabilities for the data with corrected dipole component are given in brackets.
We round to integer percentiles for probabilities $<99\%$.
The saturated values are marked with $\star$. We abbreviate the results consistent at given CL as: ``no significant detections (CL)''.
}
\centering
\begin{footnotesize}
\begin{tabular}{llllll}\hline\hline
$\mathbf{d}$\hspace{.3cm} & Reg.sk.\hspace{.3cm} & $P_{rej}^m$[\%]\hspace{.3cm} & $P_{rej}^\sigma$[\%]\hspace{.3cm} & $P_{rej}^S$[\%]\hspace{.3cm} & $P_{rej}^K$[\%]\hspace{.3cm} \\
(1) & (2) & (3) & (4) & (4) & (5) \\\hline

\incall{}& ALL&
\multicolumn{4}{c}{no significant detections ($68\%$)}\\
\incq{}& ALL &
\multicolumn{4}{c}{no significant detections ($68\%$)}\\

\incv{}
 & {\it HP} 2     & 59 (20)   & \\
 & {\it HP} 4     & 99.8 (35) & \\
 & {\it HP} 8     & 88 (16)   & \multicolumn{3}{c}{no significant detections ($68\%$)}\\
 & {\it LB} 32 8  & 99.5 (30) & \\
 & {\it LB} 64 8  & 94 (37)   & \\
 & {\it LB} 64 16 & 86 (24)   & \\
\incv{5}
 & HP 4 & 99.3 (15) & \multicolumn{3}{c}{no significant detections ($68\%$)}\\

\incw{} & ALL &
\multicolumn{4}{c}{no significant detections ($68\%$)}\\\hline
{\it QV} & ALL & $>99.9^\star$ & \\
{\it VW} & ALL & $>99.9^\star$ &\multicolumn{3}{c}{no significant detections ($95\%$)}\\
{\it QW} & ALL & $>99.9^\star$ &\\
\end{tabular}
\end{footnotesize}
\label{tab:all-mm-joint}
\end{table}

We now discuss the result of the all-\mm{s} analysis described in
Sect.~\ref{sec:all_joint}. The corresponding results are presented in
Table~\ref{tab:all-mm-joint}. We see that the data are consistent
with the simulations at $\sim 68$\% CL. 
The previously mentioned (in Sect.~\ref{sec:NGK}) tentative NG detection in the \incall{} data in 
kurtosis has indeed the largest probability of rejecting
across the scales ($61\%$) 
but it turns out to be statistically completely insignificant.

The large scale variance distribution is found to be perfectly consistent with the GRF simulations.

We find a significant anomaly in the dipole component of the V band channel (see also Sec.~\ref{sec:NGV})  
detected via distribution of means in both WMAP3 ($99.8\%$ CL) and WMAP5 ($99.3\%$ CL) data.

All of the difference maps (see Sect.~\ref{sec:diff_maps_tests}) 
show a very strong departures from foregrounds free, simulated, difference maps, most 
prominently detected in means distributions.

\subsection{The ``cold spot'' context}
\label{sec:cold_spot}
A NG anomalous kurtosis excess of a wavelet convolution coefficients has been reported (e.g.~\cite{2007ApJ...655...11C}) 
in the southern hemisphere, and was found to be associated with the 
locally cold spot (CS) in the CMB fluctuations around galactic coordinates \lb{209}{-57}.
In that work the wavelet convolution scales ranging from $\sim 6.6^\circ$ to $\sim 13.2^\circ$ in diameter were used, 
with anomaly being maximized at scales of $\sim 10^\circ$ with a rejection on grounds of Gaussianity assumption exceeding 99\% CL.
At the same time the authors note that the CS was not detected in the real space analyses.

The range of scales mentioned correspond roughly to the scales tested by the {\it HP} 4 ($\sim 14.6^\circ$) and 
{\it HP} 8 ($\sim 7.3^\circ$) \regsk{s} (see. Table~\ref{tab:regionalizations}).
\begin{figure}[!t]
\centering
\renewcommand{\figurename}{Fig}
\includegraphics[width=0.5\textwidth]{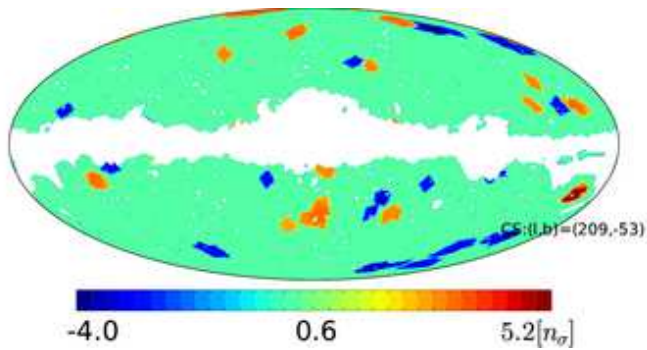}
\caption{Kurtosis n-sigma map, thresholded at $\nsigth{}=3$, combined from 100 \mm{} of the {\it HP} 8 \regsk{}.
The ``cold spot'' is  marked with ``CS''.}
\label{fig:CS_from_HP8}
\end{figure}

In Fig.~\ref{fig:CS_from_HP8} we plot the scrambled n-sigma map of kurtosis from the single-region analysis, obtained from $100$ \mm{s}
of the {\it HP} 8 \regsk{}, which most closely corresponds to the scales, at which the CS was detected. 
The deviation of $-3.6\sigma$ in the ``the cold spot'' direction, centered at \lb{209}{-53} in Fig.~\ref{fig:CS_from_HP8} is clearly found 
along with many other, locally significant, deviations. This particular CS however is not found at the same location in e.g. {\it HP} 4 \regsk{s} 
or in skewness n-sigma maps, but rather it is shifted towards smaller galactic longitudes.

As an extension to the tested scales, in this section we run a high resolution test using \regsk{} {\it HP} 16 
corresponding roughly to scales of $\sim 3.7^\circ$ and the \incall{} data.
We use 10 additionally generated \mm{s} in this resolution, and we perform the single region, joint multi-region,
and all-\mm{s} statistics. We also performed the same analysis using the filtered up to $\lmax{=40}$ is SH space, low resolution ($n_s=64$) maps
in which the spot is clearly visible.

Although we find a locally negative KE and positive excursions from expected distributions  
by $\gtrsim 3\sigma$ around the CS direction in variance, we also find similar excursions at several other directions.
The CS itself is well localized in the scrambled n-sigma maps of means with minimal value $-2.9$.

However,  none of these detections (see also Fig.~\ref{fig:all-single-reg}) hold under the scrutiny of 
the multi-region and all \mm{s} analyses (Table~\ref{tab:all-mm-joint}), that find these local anomalies to be statistically insignificant.

\subsection{Differential maps tests}
\label{sec:diff_maps_tests}
We discuss results of the QV, VW and QW difference maps tests as a simple cross-check with the CMB signal dominated maps tests, and 
a rough estimation of the residual foregrounds amplitude.
We limit the number of  \mm{s} to 10 and use $\Nsim{}=10^4$ simulations in single region analysis and $\NsimMCPDFp=10^3$
($\NsimCp = 9\cdot 10^3$) simulations in joint multi-region analysis as before.

\begin{figure}[!t]
\centering
\renewcommand{\figurename}{Fig}
\begin{tabular}{ll}
a) $\Delta_m({\rm QV})$ with uncorrected $\ell=1$&b) VW: $\ell=2$\\
\includegraphics[width=0.239\textwidth]{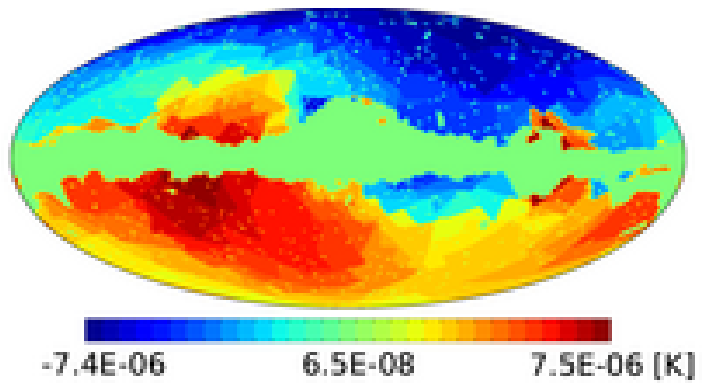}&
\includegraphics[width=0.239\textwidth]{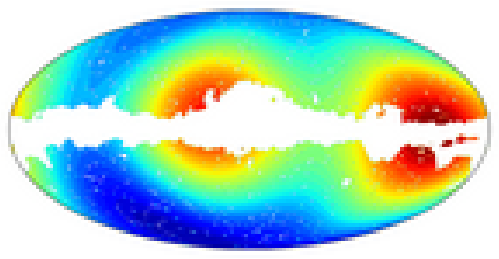}\\
\multicolumn{2}{c}{{\it HP} 2}\\
c)\makebox[0.22\textwidth][c]{$\nsig{}(QV)$} & d) \makebox[0.22\textwidth][c]{$\nsig{}(VW)$}  \\
\includegraphics[width=0.239\textwidth]{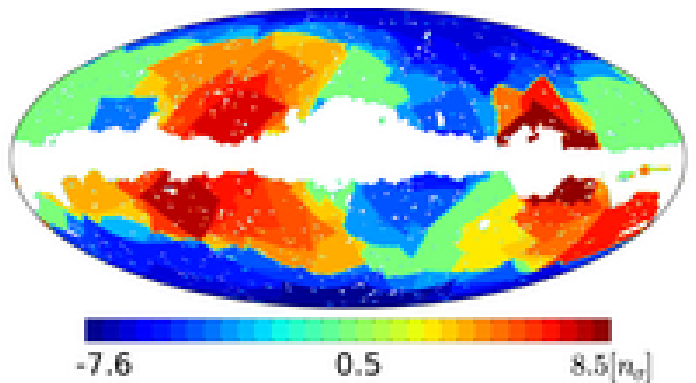}&
\includegraphics[width=0.239\textwidth]{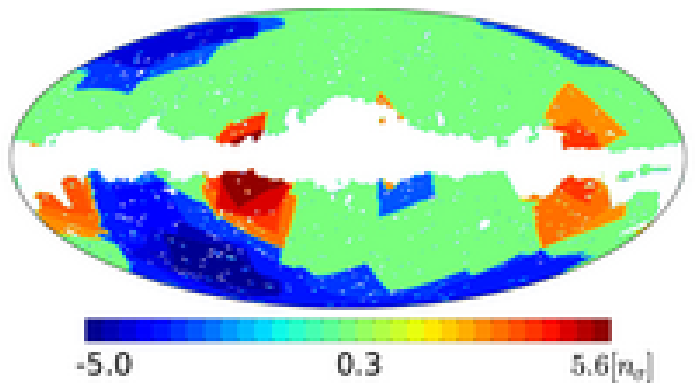}\\
\multicolumn{2}{c}{{\it HP} 4}\\
e)\makebox[0.22\textwidth][c]{$\nsig{}(QV)$} & f) \makebox[0.22\textwidth][c]{$\nsig{}(VW)$}  \\
\includegraphics[width=0.239\textwidth]{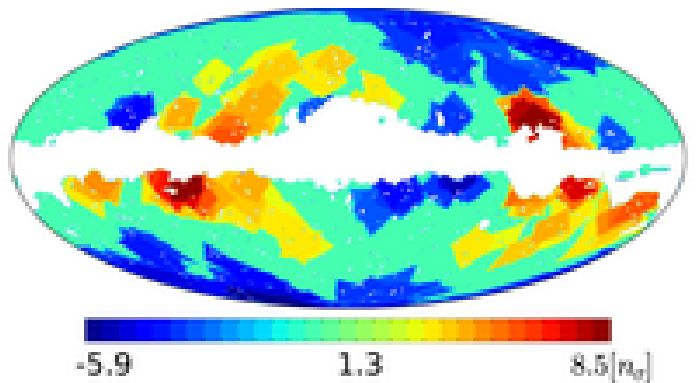}&
\includegraphics[width=0.239\textwidth]{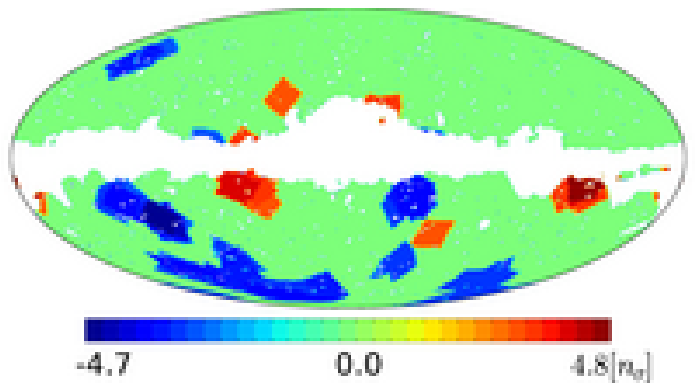}\\
\multicolumn{2}{c}{{\it HP} 8}\\
g)\makebox[0.22\textwidth][c]{ $\nsig{}(QV)$} & h)\makebox[0.22\textwidth][c]{ $\nsig{}(VW)$} \\
\includegraphics[width=0.239\textwidth]{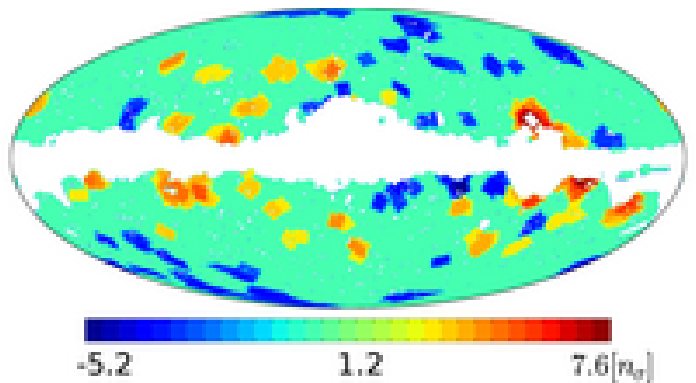}&
\includegraphics[width=0.239\textwidth]{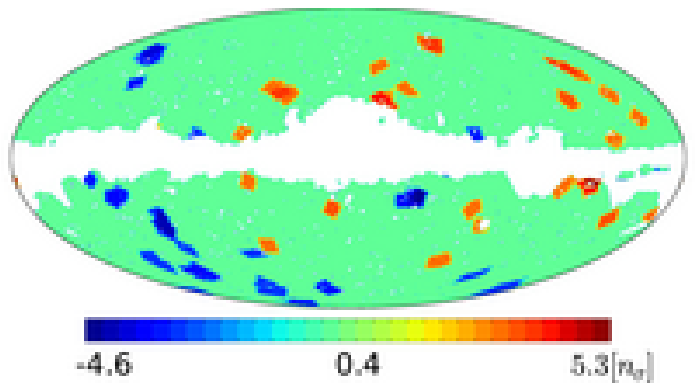}\\
\end{tabular}
\caption{Results of the single-region difference maps analysis. In panel a) the residuals ($\Delta_m({\rm QV})=m-\langle m\rangle$) 
for the difference QV data, uncorrected for the anomalous V band dipole is plotted. In panel b) the quadrupole of the VW difference map 
and in panels c) through h) for the three {\it HP} \regsk{s}, we plot the thresholded at $3\sigma$, $\nsig{}$ maps of means distributions of the differential 
datasets QV (left) and VW (right) with the anomalous dipole component removed from the data. We make these maps 
(and maps of higher MODs) publicly available at \pubdataaddr }
\label{fig:diffdeltamaps}
\end{figure}

As shown in Table~\ref{tab:all-mm-joint} the residual foregrounds are very strongly ($>99.9\%$ CL) detected in all difference maps,
due to anomalies in means distributions. 
In particular the n-sigma and difference ($\Delta=m-\langle m\rangle$) maps of means distributions in QV, most prominently exhibit
a dipole like structure oriented roughly at \lb{~260}{~60} which is close to the kinetic dipole 
direction (Fig.~\ref{fig:diffdeltamaps}a).
The VW  $\nsig{}$ map has a similar dipole structure, but with opposite orientation, which however is absent in the QW $\nsig{}$ 
map. We find this to be a consequence of the previously detected (Sect.~\ref{sec:NGV}) anomalous dipole component 
of the V band channel. Removing the dipole components from the difference maps, we have redone the three stages of the analysis, 
and although the dipole structure ceased to dominate, we still find a very high ($>99.9\%$ CL) rejection probabilities for in 
means as quoted in Table~\ref{tab:all-mm-joint}. 

In Table~\ref{tab:diff_maps} we present the amplitudes of the residual foregrounds as measured by the variance of the 
$\Delta$ difference maps of means distribution for different scales as probed by our \regsk{s}.
These remain in a good consistency with the limits given in \cite{2003ApJS..148....1B} for residual foregrounds contamination.

In Fig.~\ref{fig:diffdeltamaps}(c-h) we show the $\nsig{}$ maps with distribution of the regions in the difference maps
outstanding from simulations at significance larger than $3\sigma$ (i.e. we use $\nsigth{}=3$). Clearly, the 
close galactic plane regions are strongly detected. We note that the the KQ75 sky mask partially removes the most affected regions
around \lb{233}{-10}, \lb{259}{18} and the previously-mentioned \lb{199}{-23}.

It is interesting to note that the largest scale negative \mbox{($\nsig{}<0$)} anomalies seen in {\it HP} 2 (Fig.~\ref{fig:diffdeltamaps}c-d)
away from the galactic plane, can also be induced by the foregrounds dominating along the galactic plane (with \mbox{$\nsig{}>0$)}
due to a very strong linearity of foregrounds induced quadrupoles with strong maximums aligned along the galactic plane and 
consequently strong minimums allocated close to the poles (Fig.~\ref{fig:diffdeltamaps}b). 
Such mechanism of foregrounds-generated linearity of the quadrupoles of the difference maps
will not work in the foregrounds-free simulations, adding thereby to the observed large scale anomalies as probed via the largest
regions. This effect is considerably smaller in the higher resolution \regsk{s}.\\

In order to test the consistency of our noise simulations with the noise of  the WMAP data, 
and the approximation the uncorrelated, white noise and to constrain limits of the 
systematical uncertainties, we also performed analysis using Q12 and V12 difference maps in {\it HP} 2 \regsk{}. 
The details of this analysis is given in Appendix~\ref{app:q1q2diff_test}. 
Here, we briefly report the result that the systematical effects measured, as before, by the standard deviation
of the difference $\Delta$ maps remains at level $<1.7 {\rm \mu K}$ at the scales corresponding the {\it HP} 2 \regsk{}
i.e. $\gtrsim 30^\circ$. 
\begin{table}[!bt]
\caption{Residual foregrounds amplitudes in the cross-band difference maps.
The columns contain: (1) \regsk{}, (2) approximate angular size of the region as inferred from number of regions, (3) approximate
corresponding multipole number $\ell=180 / \Omega_{\rm reg}$, (4..6) - standard deviation of the difference maps outside \kp sky mask.}
\label{tab:diff_maps}
\centering
\begin{footnotesize}
\begin{tabular}{cccccc}\hline\hline
Reg.sk. & $\Omega_{\rm reg}$ [deg] & $\ell$ & ${\rm \sigma(QV)}$ [$\mu$K] & ${\rm \sigma(VW)}$  [$\mu$K] & ${\rm \sigma(QW)}$  [$\mu$K]\\
(1) & (2) & (3) & (4) & (5) & (6) \\\hline
{\it HP} 2 & 29.3   & 6 & 2.7 & 1.9 & 3.8\\
{\it HP} 4 & 14.6   & 12& 3.0 & 3.0 & 4.7\\
{\it LB} 32 8 & 12.7& 14& 3.9 & 3.5 & 5.1\\
{\it LB} 64 8 & 9.0 & 20& 4.7 & 4.6 & 6.0\\
{\it HP} 8 & 7.3    & 25& 4.9 & 4.9 & 6.2\\
{\it LB} 64 16& 6.3 & 29& 5.8 & 6.0 & 7.2
\end{tabular}
\end{footnotesize}
\end{table}

\section{Discussion}
\label{sec:discussion}

\subsection{Sensitivity, correlations and extensions}
Although we have shown that the statistics is rather helpless to robustly detect the  NG 
signals considered (defined in Sec.~\ref{sec:sensitivity_to_NG}) via MODs higher than the variance, 
the statistics also proved to be a sensitive and precise tool for statistical isotropy 
measurements via variance and means.
While we fail to detect the NG templates (inducing signals of rms $\sim 100{\rm \mu K}$ within spots of $\sim 10^\circ$ via skewness and kurtosis) 
in the multi-region NG analysis, the single region analysis
detects these as locally significant ($\nsig{}\gtrsim 3$). 
Instead such template can be detected in the all-\mm{s} analysis  at $>99.8\%$ CL via variance.\\

In \cite{2004ApJ...605...14E} a regional statistical analysis was performed
using a set of circular regions uniformly distributed across the
sky. Our analysis is similar in spirit but uses different statistics
and a richer sets of regions, varying both in size and shape. This
approach has been validated by the fact that we have shown that the
resulting statistical signal can be a sensitive function of the particular choice
of regions. 
This is most prominently seen in case of the reported dipole 
anomaly in the V band of the WMAP data (Fig.~\ref{fig:all-results}), where 
it is easily seen that depending on the choice of the \regsk{} 
(e.g. such as those associated with the results at the right-hand side of the top plots
in Fig.~\ref{fig:all-results}) the obviously strong anomaly can be overlooked.

It is important to mention 
the correlations between various \mm{s}. 
This correlation occurs since, although the \mm{s} sample
the data differently, in the end, the same data are being sampled.
The degree of redundancy is 
directly connected to the number of \mm{s} used in the
analysis and the magnitude of the correlation is related to the size of the regions used in a \regsk{}. 
To quantify this we carried out a simple statistical test 
only whose goal was to establish whether our test is statistically more sensitive to the change of the tested data
itself or to the change of the \mm{s} for various resolutions parameter $r$ and MOD. 

If the results from different \mm{s} were strongly correlated, then for
a constant number of DOF one would expect  the variance of the results
measured over these \mm{s} (e.g. $\chisq$ values) to be much smaller
than the variance computed while fixing a \mm{} but varying
dataset. On the other hand, if the variance of the results when
changing dataset was much smaller than the variance when changing
\mm{}, then the test would not be very sensitive to particular
features in the data, or even unstable. To test this we calculated the
$\mathit R$ statistic defined as follows: 
\begin{equation}
\mathit{R}(r,{\rm MOD}) = \frac{
  \bigl<\sigma_{\rm{sim}}(\chisq(r,{\rm MOD})/\DOFeff)\bigr>_{m} }{
  \bigl<\sigma_m(\chisq(r,{\rm MOD})/\DOFeff)\bigr>_{\rm{sim}} } 
\label{eq:R_stat}
\end{equation}
where  $\bigl<\bigr>_{\rm{sim}}$ denotes an average over simulations and $\bigl<\bigr>_{m}$ denotes an
average over \mm{} for a given simulation, and where $\DOFeff = DOF(r,{\it m})$ is the
effective number of degrees of freedom.
Measuring
this $R$ statistic using our simulations, we find that the test is
approximately equally sensitive at all resolutions and for all MODs, and that eventually it is  little more
sensitive to the change of the data under test than to the change of
the \mm{} giving $\mathit{R}$ values around $1.2$.\\

The approach with arbitrary shape of the regions (\mm{}) and their
orientation is quite flexible, and different shapes can possibly be
used for different applications independently of the enforced sky
cuts. This allows for a thorough test of the multivariate nature of
this Gaussian field. One could also consider other statistics than the
first MODs, as e.g. regional Minkowski functionals.  
Another possible extension is to apply 
a specific pre-filtering of the data in the SH space in order to expose for the test features 
dominating at particular scale. Such slicing of the data into subsets of maps according to 
some chosen ranges of multipoles could in principle significantly improve the sensitivity.
The multi-region full-sky analysis though is restricted generally to the resolutions up to which
the full covariance matrix analysis is feasible.

\section{Conclusions}
\label{sec:conclusions}
We introduce and perform a regional, real space test of statistical isotropy  and Gaussianity of the WMAP CMB data, 
using a one-point statistics.
We use a set of regions of varying size and shape (which we call \mm{}) allowing for an
original sampling of the data. For each of the regions we analyze
independently or simultaneously,  the first four moments of distribution 
of pixel temperatures (i.e. mean, variance, skewness and kurtosis).

We assess the significance of our measurements in three different
steps. First we look at each region independently. Then we consider a
joint multi-region analysis to take into account the spatial correlations between different regions. Finally
we consider an ``all-\mm{s}'' analysis to assess the overall significance
of the results obtained from different sky pixelizations.

We show that the results of such multi-region analyses strongly depend on the way in which the sky is partitioned into regions for the 
subsequent statistics and that our approach offers a richer sampling of the underlying data content.
Consequently the all-\mm{s} analysis provides a more robust results, avoiding possible biases resulting from an analysis constrained only to a single choice of \regsk{}.

We find the three-year WMAP maps well consistent with the realistic, isotropic, Gaussian Monte-Carlo simulations as probed 
by regions of angular sizes ranging from $6^\circ$ to $30^\circ$ at $68\%$ confidence level.

We report a strong, anomalous ($99.8\%$ CL) dipole ``excess'' in the V band of the three-year WMAP data and also
in the V band of the WMAP five-year data ($99.3\%$ CL) (Sect.~\ref{sec:NGV}).

We test the sensitivity of the method to detect particular CMB modulation signals defined via the scale dependent modulation amplitude parameter ($\Almax{}$) 
for the case of modulation extending up to maximal multipole number of $\lmax=40$ and $\lmax=1024$. 
We are able to reject the modulation of amplitude 
of $A_{1024}=0.114$ at $>99.9\%$ CL and find that $A_{40}=0.114$ can be statistically excluded only at $\sim 92\%$ CL 
(Sect.~\ref{sec:sensitivity_to_power_anomalies},~\ref{sec:SGsigma}).
Given the WMAP \incv{} band data, we find that the large-scale hemispherical asymmetry is not highly statistically 
significant in the three-year data ($\sim 97\%$) nor in the five-year data ($\sim 93.5\%$)  at scales $\ell \leq 40$.
Including a higher-$\ell$ multipoles only decreases the significance of hemispherical variance distribution asymmetry.

We also test the sensitivity to detect a broad range of small ($10^\circ$ in radius) locally introduced NG signals, inducing non-vanishing kurtosis (and in general skewness)
of rms amplitude $\sim 100 {\rm \mu K}$ and find that the method is able to detect these as locally significant, but the overall impact in the joint 
multi-region analysis is unnoticed by mean, skewness and kurtosis, but is strongly detected ($\sim 99.8\%$ CL) by variance distributions. We conclude that the NG foreground-like signals will be easier to detect using local variance measurements 
rather than higher moments-of-distribution. 

We also analyze our results in context of the significance of the ``cold spot'' (CS),
reported as highly anomalous at scales corresponding to $\sim 10^\circ$ in diameter.
While we notice the cold spot region as having locally anomalous, negative kurtosis-excess 
and locally increased variance (eg. Figs.~\ref{fig:all-single-reg},~\ref{fig:CS_from_HP8}),
we do not find these deviations to be globally statistically significant.

We easily detect the residual foregrounds in cross-band difference maps at average rms level $\lesssim 7{\rm \mu K}$ 
(at scales $\gtrsim 6^\circ$)
and limit the systematical uncertainties to $\lesssim 1.7{\rm \mu K}$ (at scales $\gtrsim 30^\circ$) 
as a result of the analysis of same-frequency difference maps.
These levels are consistent with the previously estimated limits.

\section*{Acknowledgements}
BL is grateful to Olivier Dor\'e for useful comments, discussions, and encouragement.
BL would like to thank Naoshi Sugiyama for his support and Boudewijn Roukema for
careful reading of the manuscript.
BL would also like to thank the anonymous referee for useful comments and suggestions.
BL acknowledges the use of the Luna cluster of the National Astronomical
Observatory (Japan), computing facilities of the Nagoya University 
and support from the Monbukagakusho scholarship. 
\par We acknowledge the use of the Legacy Archive for
Microwave Background Data Analysis (LAMBDA). 
Support for LAMBDA is provided by the NASA Office of Space Science.

\bibliography{current} 
\bibliographystyle{aaeprint}
\appendix

\section{WMAP simulations}
\label{app:simulations}
\subsection{Signal and noise pseudo $C_\ell$ tests.}
In order to assess the confidence levels we have performed Monte-Carlo
simulations of the signal ($\ell_{max}\leq 1024$ and $C_{0,1}=0$) and
inhomogeneous but uncorrelated Gaussian noise maps for all DAs, 
according to the best fit to the $\Lambda CDM$ model power spectrum,
extracted from observations \citep{2007ApJS..170..288H}. The simulations include the
WMAP window functions for each DA. The $\Nsim( = 10^4)$ full sky
simulations were generated at the \HEALPIX  resolution
${n_s=512}$, for each DA and preprocessed in the same way as the data.  

\par
\begin{figure}[!hbt]
\centering
\renewcommand{\figurename}{Fig}
\includegraphics[angle=-90,width=0.5\textwidth]{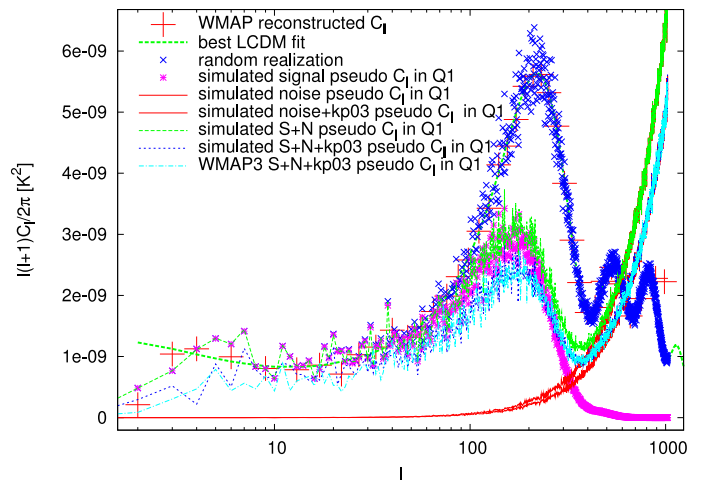}
\caption{Consistency check between the simulations and WMAP three-year observations.
The pseudo-$C_\ell$ power spectra of the WMAP (light blue/dot-dashed curve) and its simulation (the underlying blue/dashed line) in 
channel Q1 are well consistent both in \kp sky cut regime and in the noise
dominated regime. Reconstructed power spectrum of the
\cite{2007ApJS..170..288H} is plotted as big red crosses.}  
\label{fig:data_simCl}
\end{figure}

 In Fig.~\ref{fig:data_simCl} an example of simulated, and
recovered pseudo-$C_\ell$ is compared with the pseudo-$C_\ell$ of the WMAP
data from channel Q1 as a simple consistency check. Similar
simulations were performed for the remaining DAs.  

\begin{figure}[!hbt]
\centering
\renewcommand{\figurename}{Fig}
\includegraphics[width=0.4\textwidth]{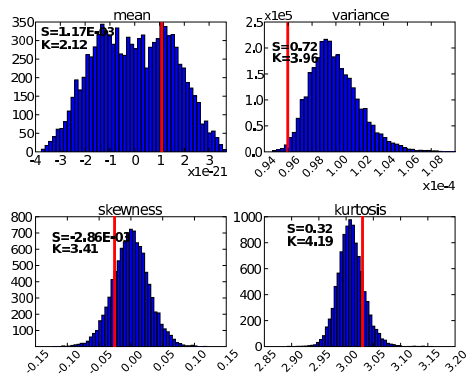}
\caption{
The distributions of means, variances, skewness and kurtosis of $\Nsim=10^4$, full sky, \incall{} simulated data realizations, calculated outside the
\kp sky mask. Skewness and kurtosis values of the distributions are also given.
Vertical bars indicate the values derived from the WMAP three-year data. 
The quantile probabilities of the mean, variance, S, and K values of the WMAP data
are $\{0.57, 0.015, 0.34, 0.35\}$ respectively -- well consistent with Gaussian, random simulations. Low probability of the 
total variance results primarily from the low quadrupole (octupole) of
the WMAP data as compared with the best fit $\Lambda CDM$ model.
Note that the distribution of the means of the simulations represents only
a numerical noise since during preprocessing all maps were shifted to zero the mean outside the \kp sky mask
($<T>\rightarrow 0$) and therefore it does not carry any important cosmological
information. The spectral analysis yields a similar results.
}
\label{fig:simshists}
\end{figure}

As a simple consistency test, we present in Fig.~\ref{fig:simshists}
statistics of the full sky WMAP third year data as compared with the
 simulations. The data are well consistent with the simulations.

\subsubsection{Tests of modulated-simulations' power spectrum}
\label{app:modulation_power_change_test}
As mentioned in Section~\ref{sec:sensitivity_to_power_anomalies} the modulation (Eq.~\ref{eq:power_mod}) alters the underlying
power spectrum of the modulated simulations and could possibly mislead us in the interpretation of the high rejection 
confidence level thresholds, reported in Table~\ref{tab:all-mm-joint-tests} for the modulated simulations, since the additional power 
could be merely a result of the inconsistency on grounds of the intrinsic power spectrum mismatch, rather than 
the regional variance distribution analysis, and violation of the statistical isotropy.

In this section we quantify this effect. We generate a set of 10 WMAP V3 simulations and modulate their CMB component
using modulations $A_{1024}(\hat {\bf{n}}=(225^\circ,-27^\circ))=\{0.114,0.2\}$. 
We next extract their \kp cut sky pseudo-power spectra up to $\lmax = 700$ i.e.
up to the scales where the noise component already strongly dominates over any possible CMB modulation signals. 
We derive the contribution
to the total variance of the map per multipole according to: 
$\sigma_\ell = \frac{2l+1}{4\pi} C_\ell$ ($\ell \in\{2,\ldots,700\}$).
We measure the degree of the consistency of the modulated simulations' power spectra with the non-modulated simulations' power spectra, 
using the unbiased estimator of the full covariance matrix $\Sigma_{ll'}={\rm Cov}(\sigma_\ell,\sigma_{\ell'})$ derived from 3000, 
analogously prepared $\sigma_\ell$ vectors extracted from
 the GRF WMAP V3 simulations, and using the corresponding Monte-Carlo probed $\chi^2$  values distribution from 223 additional simulations.\\

 We find that the average rejection thresholds for the modulations $A_{1024}=0.114$ and $A_{1024}=0.2$ are $54\%$ 
and $59\%$ respectively with the standard deviation $\sim 30\%$ in both cases. 
We therefore conclude that our results given in Table~\ref{tab:all-mm-joint-tests} cannot result
from simply alteration of the power spectra after the modulation field has been introduced. 
Similar results are obtained if the off-diagonal terms of the 
covariance matrix are neglected, which indicates that the low detection threshold does not result from the limited number of simulations and a possible 
non-convergence of the covariance  matrix.

\section{Assessing statistical significance}
\label{app:analysis}
In this appendix we give details on out statistical approaches.
For fast reference, we summarize the principal symbols used in Table:~\ref{fig:symbol-list}.
\subsection{Individual region analysis}
\label{sec:single_region}

In the case of individual regions statistics  for every $i$th region ($i\in\{1,..,\Nregrm\}$) 
 of every \mm{} $m\in\{1..\Nmm\}$ and for every MOD ($X \in \{\rm{m},\sigma,\rm{S},\rm{K}\}$) 
and every dataset ($d\in\{\incq{}, \incv{}, \incw{}, \incall{}\}$) we 
define a parameter vector $\mathbf{p}=\{r,m,d\}$ and independently
calculate the tail probability $P\bigl(\Xpi\bigr)$ 
of occurrence of $\Xpi$ 
value of the data in the $\Nsim=10^4$ simulations, probing the corresponding probability distribution functions (PDF). 
The quantities $\rm{m},\sigma,\rm{S},\rm{K}$ correspond to the first four  moments of distribution respectively.
Hence we define $P\bigl(\Xpi\bigr)$ as:
\begin{equation}
P\bigl(\Xpi\bigr) \equiv P_q\bigl( |\Xpi^{sim}-\secQpi|>|\Xpi^{data}-\secQpi| \bigr)
\label{eq:defP}
\end{equation}
where $\secQpi$ is the second quartile of the corresponding PDF, and 
$P_q$ is the quantile tail probability.

In principle, considering $N$ simulations allows us to probe a region of the underlying PDF corresponding to 
Gaussian number of ``sigmas''
\begin{equation}
\pm \nsigMC{}=\sqrt{2}\rm{erf}^{-1}(1-2/N) = |\rm{cdf_G}^{-1}(1/N)|
\label{eq:nsigMC}
\end{equation} 
where 
$\rm{cdf_G}^{-1}$ is the inverse Gaussian cumulative distribution
function (CDF). For $N=\Nsim=10^4$, as it is the case for individual
region statistics, $\nsigMC{}\approx 3.72$ corresponding to a
probability of exceeding of 0.02\%. 

\begin{figure}[!t]
\centering
\renewcommand{\figurename}{Fig}
\includegraphics[angle=-90, width=0.49\textwidth]{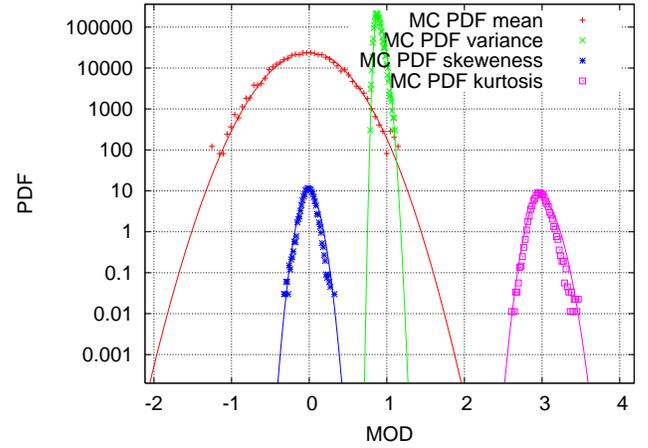}
\caption{Example of extrapolation (solid lines) used to compute the
  distribution of $10^4$ MODs from one of the regions in one of the
  \mm{s}. Mean and variance MODs values were multiplied by a factor of
  $10^4$ for the sake of clarity. Note that the Gaussian extrapolation
  (used only outside MC probed PDF region) also interpolates the data
  quite well in case of the means (red crosses).}
\label{fig:extrapolation}
\end{figure}

To derive the probability from Eq.~\ref{eq:defP}, we use linearly interpolated quantile probability within the MC probed PDF range: 
\begin{equation}
\begin{array}{lll}
  P\bigl(\Xpi\bigr) &=& Q_{lin}\bigl(\Xpi\bigr)\hspace{2mm} \rm{for} \\
&&\Xpi \in \bigl(\XMCmin,\XMCmax\bigr)
\end{array}
\label{eq:interpolation}
\end{equation}
where $Q_{lin}$ is the linearly interpolated quantile probability, and $\XMCmin$ and $\XMCmax$
are the minimal and maximal $\Xpi$ values obtained from a sample of $\Nsim$ simulations that probe the underlying $\Xpi$ PDF.
Outside the probed range
($\Xpi\in\bigl(-\infty,\XMCmin\bigr]\cup \bigl[\XMCmax,\infty\bigr)$) we extrapolate using a Gaussian distribution of the form 
\begin{equation}
\begin{array}{l}
  P\bigl(\Xpi\bigr) = \Bigl(1-\rm{erf}\bigl(\frac{n_\sigma}{\sqrt{2}}\bigr)\Bigr)\\
  \nsig{} = \Biggl\{
  \begin{array}{ccc}
    \nsigMC{} \bigl(1 + \frac{\XMCmin-\Xpi}{|\XMCmin-\secQpi|}\bigr) & ; & \Xpi \leq \XMCmin\\\\
    \nsigMC{} \bigl(1 + \frac{\Xpi-\XMCmax}{|\XMCmax-\secQpi|}\bigr) & ; & \Xpi \geq \XMCmax
  \end{array}
\end{array}
\label{eq:extrapolation}
\end{equation}
Note that the extrapolation form is connected to the MC probed PDF
range by the matching condition 
$P(\Xpi(\nsigMC{}))=P(\XMCmin) = P(\XMCmax) = 2/N$ 
-- i.e. the probability of finding a
$\Xpi$ 
value anywhere in range $(-\infty,\XMCmin]\cup [\XMCmax,\infty)$. 
An example of this extrapolation is shown in Fig.~\ref{fig:extrapolation}. This
extrapolation is obviously not validated
for MODs higher than the mean,
but we use it as a guide for very low probability events. Note that all
our results with a lower significance (roughly outside $3\sigma$ CL) are obtained modulo this approximation. \\ 

\subsection{Multi-region joint analysis}
\label{sec:multi_region}

In the multi-region analysis, we account for all
correlations between regions of a given \mm{} using an unbiased estimator of the full
covariance matrix. 
Using the same parameter vector $\mathbf{p}=\{r,m,d\}$,
we define a one column vector for each MOD ($\Xp \in\{\mathbf{\rm{m}_\mathbf{p},\sigma_\mathbf{p},S_\mathbf{p},K_\mathbf{p}}\}$) of size $\Nregrm$ such that 
$\Xp =\bigl(X_{r,m,d,i=1},...,X_{r,m,d,i=\Nregrm}\bigr)^T$ contain $\Xpi$
values for each region of a given \mm{} $m$ of given \regsk{} $r$ and
dataset $d\in\{\incq{}, \incv{}, \incw{}, \incall{}\}$. Introducing a parameter vector $\mathbf{q}=\{X,r,m,d\}$ we define
a corresponding $\chisqq$ value as:
\begin{equation}
\chisqq = (\vX^{\rm{data}}_{\mathbf{p}}-\langle\vX^{\rm{sim}}_{\mathbf{p}}\rangle)^T\iCovq (\vX^{\rm{data}}_{\mathbf{p}}-\langle\vX^{\rm{sim}}_{\mathbf{p}}\rangle)
\label{eq:chisq-joint}
\end{equation}
where the $\langle\rangle$ is the average $\vX$ from $\NsimCq$
simulations and $\iCovq$ is an unbiased estimator of the
inverse covariance matrix \citep{2007AA...464..399H} calculated from $\NsimCq$ simulations and is given by:
\begin{equation}
(\iCovq)_{ii'} =\frac{\NsimCq-\Nregrm-2}{\NsimCq-1} (\mathbf{\tilde C}^{-1}_{\mathbf{q}})_{ii'}\\
\label{eq:iCov}
\end{equation}
and $\mathbf{\tilde C}^{-1}_{\mathbf{p}}$ is the inverse covariance matrix.

\par
\begin{figure}[!t]
\centering
\renewcommand{\figurename}{Fig}
\includegraphics[angle=-90,width=0.49\textwidth]{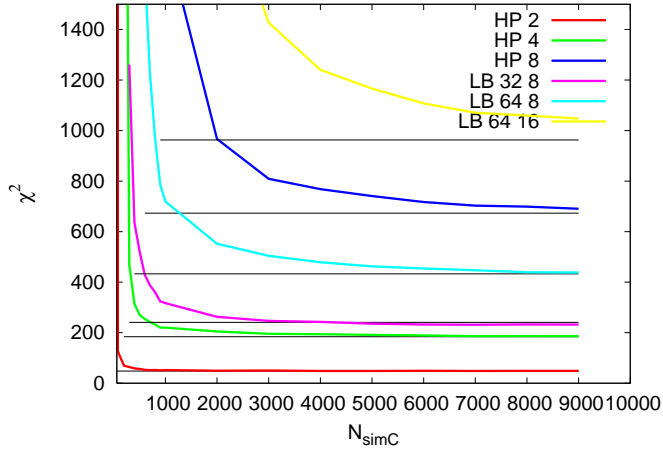}
\caption{Convergence of $\chisqq$ values as a function of the number of
  simulation used for the covariance matrix calculation for a
  given MOD. Each color corresponds to a type of \regsk{s}. The
  number of regions increases from bottom (48, for {\it HP} 2) to top (1024
  for {\it LB} 64 16) (see. Table~\ref{tab:regionalizations}). Horizontal lines
  indicate the effective numbers of degrees of freedom for that \mm{}
  (i.e. number of unmasked regions due to \kp sky mask). These $\chi^2$
  corresponds to the the first MOD only (mean) but other MODs exhibit
  similar dependence.}
\label{fig:convergence}
\end{figure}

 In Fig:~\ref{fig:convergence} we show the convergence $\chisqq$
with the number of simulations used to calculate the covariance
matrix, $\NsimCq$, for all six types of \regsk{s}. The biasing of the
$\chisqq$ values  with regards to the ideal $\chisqq(\NsimCq=\infty)$
value is caused by the limited number of available simulations. As
expected, for a given number of simulations, this bias decreases with the effective number
of DOF and as such with the \mm{} resolution. We account for this biasing by using simulations to probe the
underlying PDF of the $\chisqq$ values, instead of using theoretical
$\chisq$ distributions. In fact, it would not be valid to use
theoretical $\chisq$ PDFs,  since the distributions of individual MODs (except for the mean) are not Gaussian. 
As such, we will probe the underlying distributions using $\NsimMCPDFq
= \Nsim-\NsimCq$ simulations, for each MOD and for each \mm{}. 

In the case of the joint multi-region statistics we use
$N=\NsimMCPDFq=10^3$ simulations corresponding to $\nsigMC{}\approx
3.09$, which gives a probability of exceeding and 0.2\%. The remaining
$\NsimCq=9\cdot 10^3$ simulations are used for covariance matrix
calculation. 

Given a $(\chisqq)^{\rm data}$ value we define a corresponding joint probability as:
\begin{equation}
P\bigl(\chisqq\bigr)\equiv 1-P_q\bigl((\chisqq)^{\rm{sim}} > (\chisqq)^{\rm{data}}\bigr)
\label{eq:Pjoint}
\end{equation}
We use the same formulas for inter/extrapolation as described in Sect.~\ref{sec:single_region}
by replacing $\Xpi$ with $\chisqq$.

We note that, in fact, in case of the multi-region analysis it doesn't make any difference whether 
the derived $\chisq$ values are statistically debiased or not, since exactly the same biasing affects the simulated 
$\chisq$ values used to probe the underlying PDF. We also find that the convergence
of such derived probabilities is actually much better than one could infer when looking at the worst case of {\it LB} 64 16 
presented in Fig.~\ref{fig:convergence}. We estimate that the derived probabilities 
converge to their true values within $\sim 10\%$ of that value in this worst case of {\it LB} 64 16 \regsk{}, while in case 
of the {\it HP} 2 the convergence of the derived probabilities is $\lesssim 2\%$.\\
The statistical debasing used for the $\chisq$ values matters however in our third-stage analysis, i.e. in 
case of the all-\mm{s} analysis.

\begin{table}[!t]
\caption{List of principal acronyms used in this section and in the main body of the paper, briefly summarized for 
quick reference. \mbox{$\star$ - indicates ``unless specified otherwise''}.}
\centering
\begin{footnotesize}
\begin{tabular}{ll}\hline\hline
Symbol & explanation\\\hline
MOD & moment of distribution\\
data & upper index to indicate a measurement on data\\
sim & upper index to indicate a measurement on simulations\\
$\Nmm$ & total number of \mm{s} in a given\\& \regsk{} ($\Nmm=100^\star$)\\
$m$ & \mm{} index number $m \in \{1,\ldots,\Nmm\}$\\
$r$ & \regsk{} resolution parameter \\& $r\in \{48,192,256,512,768,1024\}$ (Tab.~\ref{tab:regionalizations})\\
X & MOD: $X \in \{\rm{m},\sigma,\rm{S},\rm{K}\}$, \\& mean, variance, skewness, kurtosis respectively\\
$d$ & dataset: $d\in\{\incq{}, \incv{}, \incw{}, \incall{}\}$\\
$\Nregrm$ & number of regions in $m$'th \mm{}\\& of the \regsk{} $r$\\
$i$ & region index within a \mm{}\\
$\mathbf{p}$ & parameter vector: $\mathbf{p}=\{r,m,d\}$\\
$\mathbf{q}$ & parameter vector: $\mathbf{q}=\{X,r,m,d\}$\\
$\Xpi$ & value of a MOD for i'th region of $m$'th \mm{}\\& of $r$'th \regsk{} in a dataset  $d$\\
$\Nsim$ & total number of simulations; \\&number of simulations used in the single-region analysis \\
$P\bigl(\Xpi\bigr)$& probability corresponding to $\Xpi$ derived from $\Nsim$ \\& simulations (Eq.~\ref{eq:defP})\\
$\nsigMC{}$ & defined in Eq.~\ref{eq:nsigMC}\\
$\Xp$ & vector of MOD values for for a given value of parameter $\mathbf{p}$\\
$\NsimCq$ & number of GRF simulations used to derive the covariance\\&  matrix in multi-region analysis ($\NsimCq=9\,000^\star$)\\
$\NsimMCPDFq$ & number of GRF simulations used to probe the distribution of \\&  $\chisqq$ values in multi-region analysis ($\NsimMCPDFq=1\,000^\star$) \\
$\chisqq$ & $\chisq$ value for the corresponding $\vX^{\rm{data}}_{\mathbf{p}}$ (Eq.~\ref{eq:chisq-joint})\\
$P\bigl(\chisqq\bigr)$ & probability corresponding to $\chisqq$ (Eq.~\ref{eq:Pjoint})
\end{tabular}
\end{footnotesize}
\label{fig:symbol-list}
\end{table}

\subsection{All {\it Multi-masks} analysis}
\label{sec:all_joint}
There is no a unique way to generalize from the results of the multi-region analysis.
Although the most straightforward way would be 
to compute the inverse covariance matrix between all the MODs and for all regions of all \mm{s}, 
this turns out to be computationally not feasible.

Note the fact that the $\chisqq$ values are partially correlated
with each other, since they result from a different sampling of the same dataset. 
However the degree of correlation strongly depends on the 
particular \mm{} properties and resolutions. In particular the
inter-\mm{} correlations are largest in the lowest resolution
\mm{s}. The smaller correlations between various \mm{s}, the more
additional information the \mm{} analysis explores about the
dataset. Large inter-\mm{} correlations indicate that not much new
information is gained making it unnecessary to process large number of
\mm{s}.\\ 

In the following, in order the integrate all the information probed by different \mm{s}
we calculate the cumulative probability of rejecting the GRF hypothesis 
using the median distribution $\overline{\varphi}(\chisqqMC)$ of all processed $\chisqqMC$ distributions and 
the median $\overline{\chisqq}$ value of the data.
Therefore we calculate the distribution and $\chisqq$ value as:
\begin{equation}
\begin{array}{lll}
\overline{\varphi}(\chisqqMC) &=& \langle\varphi(\chisqqMC/\DOFeff)\rangle_m\\\\
\overline{\chisqq} &=& \langle\chisqq/\DOFeff\rangle_m
\end{array}
\label{eq:chisq-jointall}
\end{equation}
where 
$\langle\rangle_{\rm{m}}$ is the average over all \mm{s}, 
$\DOFeff = {\rm DOF}(r, m)$ is the
effective number of degrees of freedom which is a function of
resolution $r$, of the \regsk{} and of a particular \mm{} due to
interplay with the sky cut.

We define the joint cumulative
probability of rejecting the GRF hypothesis of the WMAP CMB data via
inconsistency with our simulations as a function of $\mathbf{q}$
analogically as in Eq.~\ref{eq:Pjoint}, and we use the same extrapolation and interpolation 
formula as in case of multi-region analysis.

\section{Noise simulations tests}
\label{app:q1q2diff_test}

Difference maps obtained from observations in nearly the same frequency, and with nearly the same beams profile provide 
a good opportunity to measure the statistical properties of the instrumental noise. 

We have performed a reduced $\chi^2$ tests, directly in pixel domain, of the difference maps obtained from different 
channels of the WMAP data (Q12, V12, QV and Q1V1) at the Healpix resolution $n_s=512$, and compared with results 
of the same tests performed at a low Healpix resolution $n_s=4$. 

Since the covariance matrix of the noise realizations is well diagonal a single variate Gaussian statistics was assumed,
and reduced $\chi^2$ distributions used.
The Q12 and V12 yielded a well consistency with the simulations at both resolutions. The QV and Q1V1 difference data
however, turns out to be more troublesome.  Whereas there is a good consistency at high resolution, the low resolution
reduced $\chisq$ tests show significant discrepancy yielding a ``probability of rejecting'' $P=0.999963$ in case of QV map 
and $P=0.998$ in case of Q1V1 map. This result is also discussed in Sect.~\ref{sec:diff_maps_tests} in light of the 
anomalous dipole component of the V band map.

We also performed a single-region, joint multi-region, and all \mm{s} analyses on the Q12 difference map, 
using a subset of 10 selected \mm{s} of the {\it HP} 2 \regsk{}. 
Since the low resolution analysis yields a quick convergence (Fig.~\ref{fig:convergence} in
Sect.~\ref{sec:multi_region}) a small number of 500 simulations were generated and half of them used for covariance matrix
calculation, and the other half was used for probing distributions of the $\chisq$ values.

We found strong anomalies in the distribution of means (of which joint probability is well extrapolated using 
Eq.~\ref{eq:extrapolation} outside the MC probed range (Fig.~\ref{fig:extrapolation})). The variance of the scrambled $\Delta$ maps show
that the rms amplitude of the differences is limited to the $1.7 {\rm \mu K}$ at these scales which is consistent with the 
limits to the residual systematical uncertainties in Q1 and Q2 channels of the WMAP \citep{2003ApJS..148...63H} at these scales.
The constraint includes not only the systematical uncertainties but also possible differences due to uncorrelated white noise
used in our simulations, which in principle in the regional statistics do not average out in the same way as the pre-whitened 
1/f pink noise of the WMAP data.

A difference $\Delta$ map of the variances can also serve as a rough estimate of the level of local systematical effects.
Anomalies in the scrambled map are indeed found, with strongest deviations concentrated in parts of regions adjacent to the 
Galactic Center, with extreme values $<3 {\rm \mu K}$. 
However a close orientation of the regions to the Galactic Center
is more likely a hint on the residual foregrounds contamination, due to 
slight differences in the effective frequency of the Q1 and Q2 differential assemblies, as well as in the beam profiles,
rather than a manifestation of a systematical effects. Due to this leakege the limits to the aforementioned systematical effects at the level of $1.7 {\rm \mu K}$  
should be considered as an upper limits.

\end{document}